\documentclass[pra,twocolumn,longbibliography]{revtex4-1}
\usepackage[dvips]{graphicx}%
\usepackage{bm,color}
\usepackage{amsmath,amssymb}
\newcommand{\braket}[2]{\langle #1 | #2 \rangle}
\newcommand{\ketbra}[2]{\ket{#1}\!\bra{#2}} 
\newcommand{\ket}[1]{\left |  #1 \right \rangle}
\newcommand{\bra}[1]{ \left \langle #1  \right |}

\def \tr{{\textrm {Tr}}}
\begin{document}
\title{Schmidt-number benchmarks for continuous-variable quantum devices} 

\author{Ryo Namiki}
\affiliation{Institute for Quantum Computing  and Department of Physics and Astronomy,
University of Waterloo, Waterloo, Ontario, N2L 3G1, Canada}

\date{\today}
\begin{abstract} 
 We present quantum fidelity benchmarks for continuous-variable (CV) quantum devices to outperform quantum channels which can transmit at most $k$-dimensional coherences for positive integers~$k$. 
We determine an upper bound of an average fidelity over Gaussian distributed coherent states for quantum channels whose Schmidt class is $k$. 
This settles fundamental fidelity steps where the known classical limit and quantum limit correspond to  the two endpoints of $k=1$ and $k= \infty $, respectively.     
It turns out that the average fidelity is useful to verify to what extent an experimental CV gate can transmit a high dimensional coherence. The result  is further extended to be applicable to general quantum operations or stochastic quantum channels.  While the fidelity is often associated with heterodyne measurements in quantum optics, we can also obtain similar criteria  based on  quadrature deviations determined via homodyne measurements.    
\end{abstract} 

\maketitle


\section{Introduction}
It is a fundamental question how to generate and characterize
 higher dimensional entanglement on  quantum systems \cite{Horo09,Pan12}.  
A central tool to  identify higher dimensional entanglement 
is  the Schmidt number
 \cite{Ter20}. It is a convex roof extension of the Schmidt rank for pure bipartite quantum states, i.e., the rank of marginal density operators. 
 A quantum state of a Schmidt-class $k$ implies the state can be expressed as a  mixture of pure states whose Schmidt rank is at most $k$ for $k= 1,2,3, \cdots$. 
 On the level of quantum channels,  the Schmidt-class $k$ implies that there exists a Kraus representation in which the maximum rank of Kraus operators is at most $k$  \cite{Hua06,Chru06,Namiki13a}.  A channel of  Schmidt-class $k$ is also referred to as  $k$-partially entanglement breaking ($k$-PEB) since it  represents an important class of completely positive (CP) maps called entanglement breaking in the case of  $k=1$ \cite{Horo03a,Hol08}.   The notion of the Schmidt number  tells us a precise meaning of the dimensionality in quantum object, and enables us to demonstrate multi-level coherences of quantum gates \cite{Namiki12R} as well as to verify higher order entanglement in practical conditions \cite{Sanpera01,Tokunaga06,Tokunaga08,Inoue09,Li10,Sperl11,Namiki12L,Shahandeh2013,Guti14}.

Quantum continuous-variable (CV) systems play a central role in quantum optics and experimental quantum information science \cite{Bra05,Hamm10,Weed12}. They are described by  a set of bosonic field operators and  capable of simulating any finite dimensional quantum information process in principle. However,  their versatility could be limited due to various imperfections in experiments, and  is not necessarily  accessible in the original form of  the theoretical model. 
Hence, it is natural to ask to what extent  a given CV system is capable of  simulating a higher dimensional quantum information process in practice. Notably, a verification scheme of higher dimensional entanglement of CV quantum states has been proposed \cite{Sperl11,Shahandeh2013}. However, it has little been studied how to verify  higher dimensional gate coherences in CV quantum gates.

A practical measure to show a basic performance of  CV gates   \cite{Furusawa98,julsgaard04a,Lob09} is an average fidelity over an input ensemble of Gaussian distributed coherent states ~\cite{Bra00,Ham05,Namiki07,Takano08}.  As an ultimate limitation of gate performance, the quantum limit fidelity was determined in Refs.~\cite{Namiki11R,Chir13}. On the other hand, the entanglement-breaking limit fidelity, which is normally referred to as  the classical limit fidelity,  was determined in  Ref.~\cite{Ham05,Namiki07,Namiki11,Chir13,Yang14,Namiki1503},  
and established a practical quantum benchmark for CV gates.   
 Similarly to other quantum benchmarks \cite{Fuc03,Namiki08,Has08,Has09,Owari08,Namiki-Azuma13x}, the fidelity-based benchmark enables us  to eliminate the possibility that the process is described by  entanglement-breaking maps when the experimental fidelity is higher than the classical limit. Therefore, it can ensure the existence of the coherence in the lowest order of  $k=2$,  but  could not provide evidence of  substantially higher order coherences expected in CV gates.

  Typically, we consider a higher fidelity implies a better gate performance, and 
  it is likely that a higher fidelity suggests a higher Schmidt number and a higher order coherence.  Therefore,  an essential question is  how high the fidelity  need to be in order to outperform  a wider class of  lower dimensional processes which belong to the Schmidt class of a given Schmidt number $k$.   
     Although the known Schmidt-number benchmarks   
     \cite{Namiki12R,Namiki1503}  could be usable in general, it is crucial to observe the gate performance using more accessible quantum optical measurements  \cite{Namiki1502}.  
There are other possibilities to assess the gate coherence  quantitatively by using different measures of entanglement \cite{Kil11,Kil12,Khan13,Namiki1502}.

In this paper, we present  Schmidt-number benchmarks for CV quantum devices based on an average fidelity over Gaussian distributed coherent states. We show an upper bound of the average fidelity achieved by $k$-PEB channels for any given positive integer $k$.  It gives  general fidelity steps that reproduce the classical limit and  quantum limit for  $k=1$ and $k = \infty $, respectively. Surpassing the $k$-th limit assesses the existence of $(k+1)$-dimension coherences on quantum channels and operations. 
 We also provide a simple conjectural form of the tight $k$-th limit.  This conjectured bound  is partly achieved by a quantum channel with Schmidt-class $k$ and fully achievable by a probabilistic gate with Schmidt-class $k$, for every $k$.  Furthermore, the fidelity bound is utilized to provide  a different form for Schmidt-number benchmarks testable by using homodyne measurements.

The remainder of this paper is organized as follows. 
In Sec.~II,
we  define the Schmidt-class-$k$ limit of the average fidelity for Gaussian distributed coherent states, and show how to find an upper bound.  
In Sec.~III,
we extend the resultant fidelity-based benchmarks for probabilistic  quantum channels. 
In Sec.~IV, 
we show a lower bound of an average quantum noise of canonical quadrature variables to outperform $k$-PEB operations as well as $k$-PEB channels. 
In Sec.~V, we conclude this paper with remarks.

\section{Schmidt-class-$k$ fidelity limits for quantum channels}
\subsection{Ansatz}
We consider transmission of coherent states 
  $\ket{ \alpha}:=D(\alpha) \ket{0} = e^{-|\alpha |^2 /2} \sum_{n=0}^\infty \alpha ^n \ket{ n} /\sqrt{n!}$
   through a quantum channel $\mathcal E$.   
Let us consider a transformation task on coherent states $\{| \alpha\rangle \} \to \{\ket{\sqrt \eta \alpha }\} $ with $\eta > 0$,  and define the average fidelity for Gaussian distributed coherent states as \cite{Bra00,Ham05,Namiki07}  
\begin{align}
F_{\eta, \lambda} (\mathcal E)  :=  
 \int p_\lambda( \alpha )\bra{\sqrt \eta \alpha }\mathcal E \left(| \alpha \rangle \langle \alpha| \right) \ket{\sqrt\eta \alpha} d^2 \alpha , \label{eq1} 
\end{align}
where  $p_\lambda( \alpha ) =  \frac{\lambda }{\pi} \exp (- \lambda |\alpha |^2 )$ with  $\lambda >0 $.  
We define the Schmidt-class-$k$  fidelity limit of quantum channels by 
\begin{align}
F^{(k)}(\eta, \lambda) 
  :=& \max_{\mathcal E \in \mathcal O_k } F_{\eta, \lambda} (\mathcal  E) , \label{defk} 
\end{align}
  where $ \mathcal O_k$ is the set of $k$-PEB channels \cite{Hua06,Chru06,Namiki13a}. This set can be defined in terms of Kraus operators $ \sum_i  K_i ^\dagger  K_i = \openone  $ as 
  \begin{align}
 \mathcal O_k= \left\{ \mathcal E  \! \ \Big| \ \! \mathcal E (\rho ) = \sum_i K_i \rho K_i ^\dagger    \wedge    \forall i,\  \textrm{rank} (K_i) \le  k  \right\}.   \label{kpeb}  
\end{align}
Note that $\mathcal O_1$ represents the set of entanglement-breaking channels and $F^{(1)}$ corresponds to the classical limit fidelity \cite{Ham05,Namiki07}. Note also that $\mathcal O_{\infty}$ forms the  set of whole trace-preserving CP maps and $F^{(\infty)}$ corresponds to the quantum limit fidelity \cite{Namiki11R}.  Therefore,  $F^{(k)}$ of Eq.~(\ref{defk}) presents unified  fidelity steps which include the classical limit and quantum limit as the two endpoints, $k=1$ and $k= \infty$. 
Our main goal  is to find a non-trivial upper bound of $F^{(k)}$ for every integer $k \in [2, \infty)$.

Note 
that  there is a general definition of  PEB channels for CV systems \cite{Shirokov13}.  How to  incorporate this general definition into  our  approach is  beyond the scope of this paper.  


\subsection{Fidelity bounds} \label{bounding}
In order to find an upper bound of the fidelity $F^{(k)}$, we introduce a pair of  two-mode states \cite{Namiki11,Namiki11R} 
 as 
\begin{align}
\rho_{\mathcal E} : =& \mathcal E \otimes I \left( \ketbra{\psi_\xi}{\psi_\xi} \right ), \label{cms} \\
M: =& \int p_s(\alpha) \ketbra{ \alpha  }{ \alpha  } \otimes \ketbra{\kappa  \alpha ^*  }{\kappa \alpha ^* }  d^2\alpha ,
\end{align}
where     $I$ denotes the identity process, $\ket{\psi_\xi}= \sqrt{1-\xi ^2} \sum_{n=0}^\infty \xi^n\ket{n}\ket{n}$ is a two-mode squeezed state  with $\xi \in (0,1) $, and we assume $s, \kappa  > 0 $.
Using  the relation $\braket{\alpha }{\psi_\xi}  =  \sqrt{1-\xi ^2} e^{-(1-\xi ^2 )|\alpha | ^2 /2} \ket{\xi  \alpha^* }$  
 we can find a state-based representation of  
 the fidelity  in Eq.~\eqref{eq1} as 
\begin{align}
 F_{1/N, \tau / N} (\mathcal E )   &=  \frac{s+ (1-\xi ^2)\kappa ^2 }{s(1-\xi^2)}  \tr  (\rho_{\mathcal E} M ),  
 \label{jj}
\end{align} where the parameters $(N,\tau)$ in the fidelity function are determined by   
\begin{align} 
\tau &=  s+ (1-\xi^2)\kappa ^2, \   N  =  \kappa ^2 \xi^2  .  \label{n} \end{align}
 From Eqs.~(\ref{defk})~and~(\ref{jj}) we have  
\begin{align}
 F ^{(k)} (1/N,   \tau /N) 
 & =   \frac{s+ (1-\xi ^2)\kappa ^2 }{s(1-\xi^2)} \max_{\mathcal E \in \mathcal O_k }  \tr  (\rho_{\mathcal E} M )  
   . \label{main} 
\end{align}
 If $\mathcal E$ is a $k$-PEB channel, $\rho_{\mathcal E} = \mathcal E \otimes I \left( \ketbra{\psi_\xi}{\psi_\xi} \right )$ is a state of Schmidt-class $k$. This implies that the term $ \max_{\mathcal E \in \mathcal O_k } \tr (\rho_{\mathcal E} M )$ in Eq.~(\ref{main}) can be  upper bounded
  as    
\begin{align}
 \max_{\mathcal E \in \mathcal O_k}   \tr  (\rho_{\mathcal E} M )   \le& \max_{ \phi \in S_k} \bra{\phi}M \ket{\phi},  \label{ok} 
\end{align}
where $ S_k$ denotes the set of pure states whose Schmidt rank is $k$ or less than $k$. 

 To proceed, we use the fact that $M$ is invariant under the collective rotation $e^{i \theta (\hat n_b - \hat n _a)}$. Here, $\hat n_a \ ( \hat  n_b$) stands for  the number operator of the first  (second) mode. This implies that $M$ can be decomposed into the direct-sum form associated with the eigenspaces of the relative photon-number operator $\hat n_b- \hat n_a = \sum_J J \openone^{(J)}$ as  
 \begin{align}  M  =  \sum_{J=-\infty}^{\infty}  \openone^{(J)} M  \openone^{(J)} =: \bigoplus_{J=-\infty }^\infty M^{(J)},  \end{align} 
 where the identities of the orthogonal subspaces 
  can be written
  as $ \openone^{(J)}= \sum_{n=0}^ \infty \ketbra{e_n^{(J)}}{e_n^{(J)}}$ with $\ket{e_n^{(J)}}:= \ket{n}\ket{n+J} $ for $J\ge0 $ and $\ket{e_n^{(J)}}:= \ket{n-J}\ket{n} $ for $J<0 $. As a consequence, an explicit form for $M^{(J)}$ is given by 
\begin{align} 
M^{(J)}  =   \frac{s }{(1+s+ \kappa ^2) }   \sum_{n, m=0}^{\infty} \gamma_{n,m}^{(J)} \ketbra{e_n^{(J)}}{e_m^{(J)}}, \label{mmmm}
\end{align} where we define
\begin{align}
  \gamma_{n,m}^{(J)}  :=   & \frac{ (n+m+|J|)!  \cdot \kappa ^J x^  {n+m+|J|}  }{\sqrt{n! m!( n+|J|)! (m+|J|)!}}   
  \label{gammanm}
\end{align} and
 \begin{align}
x   =    \frac{\kappa }{1+s+ \kappa ^2} 
 \le \frac{1}{2}. \label{definitionx}
 \end{align}
From this decomposition and theorem 2 of Ref.~\cite{Sperl11}, we can see that a Schmidt-number $k$ vector $\ket{\phi} = \sum a_{n} | e_{n}^{(J)}\rangle$ in support of  $M^{(J)}$  solves the Schmidt-number-eigenvalue problem of $M$. 
 This implies that an upper bound is given by comparing the maximum on each subspace:    
\begin{align}
\max_{ \psi \in S_k} \bra{\phi}M \ket{\phi}  
& =    \max_{J}\max_{ \phi \in S_k}  \tr ( M^{(J)} \ketbra{\phi}{\phi} ). \label{ok2}
\end{align} 

Now, concatenating Eqs.~(\ref{main}, \ref{ok}, \ref{mmmm},  \ref{ok2}) and taking the limit  $s \to 0 $  with the help of Eq.~\eqref{n} we  obtain 
\begin{align}
&F^{(k)}(1/N,   \tau /N) \le \frac{N+\tau }{1+N+\tau} \max_J \max_{\phi \in S_k} \bra{\phi}A^{(J)} \ket{\phi}, &
\end{align}
where \begin{align}
A^{(J)}:=  \sum_{n, m=0}^{\infty} \gamma_{n,m}^{(J)} \ketbra{e_n^{(J)}}{e_m^{(J)}}  \label{Amat}
\end{align} and 
$\gamma^{(J)}$ is given by Eq.~(\ref{gammanm}) with $\kappa = \sqrt{N+ \tau } $ and $x = \sqrt{N+ \tau }/ (1+ N+ \tau )$. Note that,  for  $\kappa \ge 1 $ ($ \kappa < 1 $),   the  optimization over $J \ge 0$  $( J  \le 0 ) $ is sufficient due to the relation $\kappa ^{-J} \gamma_{n,m}^{(J)}= \kappa ^{J} \gamma_{n,m}^{(-J)}$ or equivalently  $\kappa ^{-J} A^{(J)}= \kappa ^{J} A^{(-J)}$.

Since $A^{(J)}$ of Eq.~(\ref{Amat}) is essentially the same form as $L$ of Eq.~(63) in Ref.~\cite{Sperl11}, 
we can evaluate $\max_{ \phi \in S_k} \bra{\phi}A^{(J)} \ket{\phi}$ 
 by the maximal eigenvalue of all $k \times k$-principal submatrices of $A^{(J)}$. This enables us to determine an upper bound of $F^{(k)}$ as follows.   Let us write 
 a $k \times k$ principal submatrix of $A^{(J)}$ by  \begin{align}
  A^{(J)}_{\vec n } := \sum_{n,m \in   \vec n } \ketbra{e_{n}^{(J)}}{e_{n}^{(J)}} A^{(J)} \ketbra{e_{m}^{(J)}}{e_{m}^{(J)}} \label{defA}
\end{align}
where  $ \vec n=  {\{n_1,n_2, \cdots, n_k\}}$ is a set of non-negative integers in increasing order, $n_l < n_{l'}$ with $l < l' $, 
 and the number of elements is denoted by $| \vec n | =k$.  Then, we can formally express the fidelity bound as 
\begin{align}
F^{(k)}(1/N,   \tau /N) \le \frac{N+\tau }{1+N+\tau}  \max_J \max_{|\vec n|=k} \|  A^{(J)} _{\vec n } \| =: U_k ,  \label{Obound}
\end{align}
where $\| \cdot \|$ denotes the  maximum eigenvalue. 

The right-hand-side formula of Eq.~\eqref{Obound} still involves optimizations over the integer $J$ and the choice of the $k$-tuple $\vec n $.
Fortunately, we can find the maximum by checking a finite set of finite-size matrices once the parameters $(k,N,\tau )$ are fixed. This is because $\{ A^{(J)}\}_J$ is essentially equivalent to the density matrix for the  Gaussian state $M $ in the number basis, and the contribution involving sufficiently large photon-number elements is negligible.  A practical process to determine  the maximum is given in  Appendices. 
Eventually, we can find the maximum by filtering out the submatrices whose maximal eigenvalue is smaller than that of another submatrix.  In Appendix~\ref{APB},  the optimal set $\vec n$ is identified  for a couple of smaller $k$ in the case of $J=0$. Appendix~\ref{APC} generalizes the approach presented in Appendix~\ref{APB}, and gives a systematic process
 to determine the maximum over general $(J , \vec n) $ for any given integer $k \in [1, \infty) $. 

\subsection{Numerical results and application}

\begin{figure}[t]
\includegraphics[width= \linewidth]{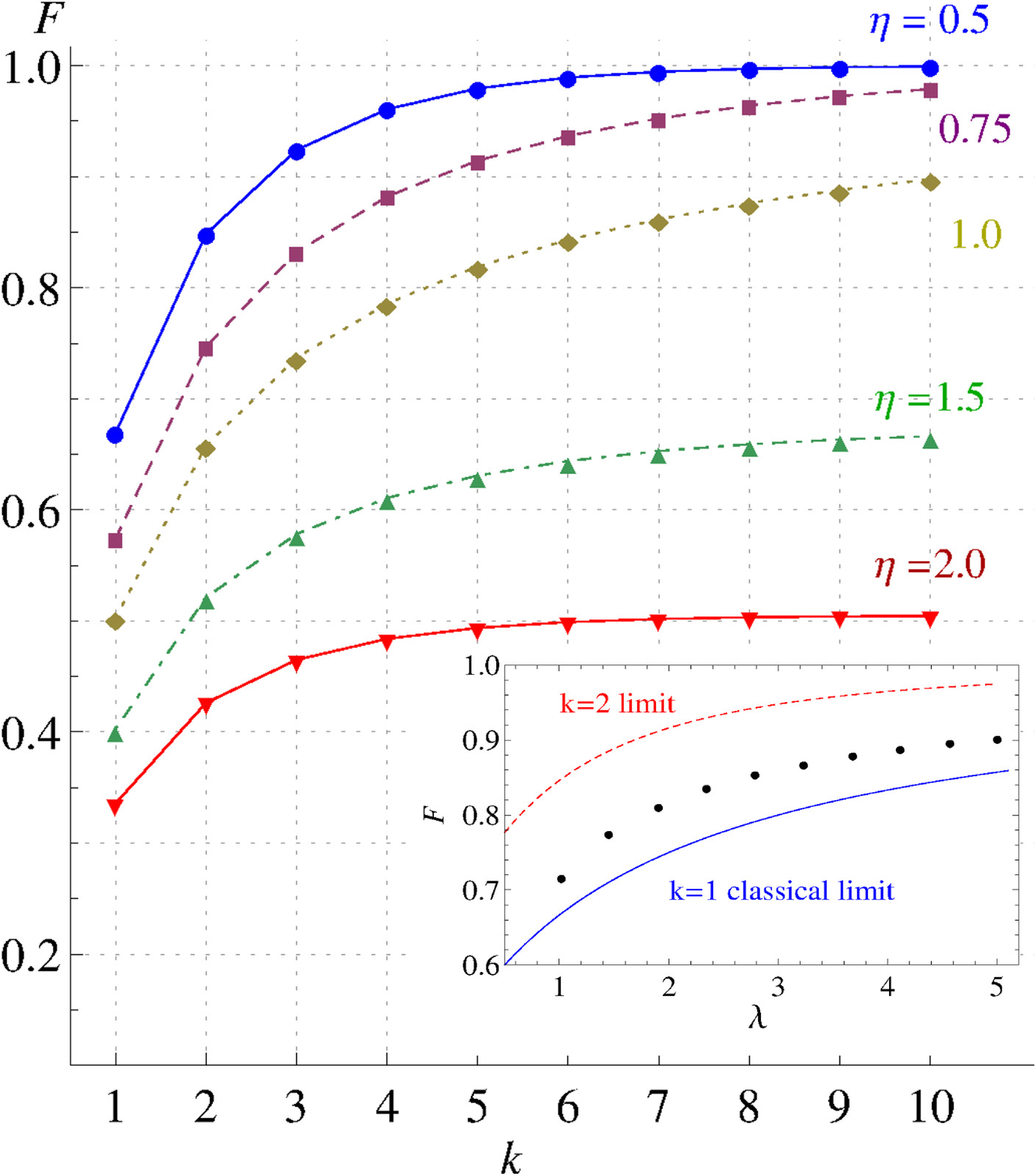}
  \caption{Our upper bound of the Schmidt-class-$k$ fidelity 
   $F^{(k)}(\eta, \lambda)$   [$U_k$ of Eq.~\eqref{Obound}] for 
  $\lambda=0.01 $ and   $\eta \in \{0.5, 0.75, 1.0, 1.5, 2.0 \}$. 
  If an experimental fidelity $F_{\eta, \lambda} (\mathcal E)$ stays above the $k$-th bound, the experimental CV~gate $\mathcal E$  cannot be described by a $k$-PEB channel. This certificates an existence  of the $(k+1)$-th coherence in the CV~gate  $\mathcal E$. The classical limit fidelity of the  fundamental quantum benchmark corresponds to the bound of $k=1$.   
   In the inset, the classical limit fidelity ($k=1$) and the fidelity bound for $k=2$  due to  right-hand side of Eq.~(\ref{conjecture}) are shown as a function of the Gaussian inverse width $\lambda$ for the case of  unit-gain condition $\eta =1 $. The dots represent the average fidelity $F_{1, \lambda} $ given  in Fig.~5 of 
    Ref.~\cite{Lob09}.
  This experimental fidelity is not high enough to give evidence to outperform an arbitrary qubit gate with regard to our criterion. 
   }  \label{fig:case10.eps}
\end{figure}

Based on the method described in Appendix~\ref{APC},  
we can numerically determine the upper bound  of $F^{(k)}$ in Eq.~\eqref{Obound}. 
Figure \ref{fig:case10.eps} shows our bound of $F^{(k)}(\eta, \lambda)$ for $k= \{1,2, \cdots, 10 \} $ and $\eta \in  \{0.5, 0.75, 1.0, 1.5, 2.0 \}$ with  $\lambda = 0.01$. For each pair of the parameters $\{\eta, \lambda\}$, surpassing the bound of  $k$ implies that the channel $\mathcal E $ outperforms $k$-PEB channels of Eq.~(\ref{kpeb}), and is capable of transmitting entanglement of Schmidt-rank $k+1$. It  certifies the quantum coherence unachievable by any teleportation-based quantum gate employing  entanglement of Schmidt-class $k$ \cite{Namiki13a}.  The fidelity steps  agree with our intuition that a higher fidelity means an existence of  stronger entanglement in terms of the Schmidt number, and would be widely useful to evaluate the performance of  CV quantum gates.

 Lobino et al., \cite{Lob09} 
  showed an experimental average fidelity as a function of $\lambda$ for unit gain $\eta=1$. In  the inset of Fig.~\ref{fig:case10.eps} we 
   find that the experimental fidelities are located in  between the lines $k=1$ and $k=2$, and  not high enough to demonstrate $k=3$ or higher dimensional coherences.  
This suggests that CV experiments are rather behind demonstrating genuinely higher dimensional coherences compared with experiments for multi-qubit channels \cite{Namiki12R}.  It might be worth noting that the current fidelity record 83\% for an experiment of a unit-gain teleportation protocol is a fidelity for an input of the vacuum state \cite{Yukawa08,Pira15np}. This corresponds to the case of  $\lambda = \infty$ in our footing, and is useless for a verification of the multi-level coherence. %




\subsection{
Conjecture and attainability} \label{subseccunjecture}
From the numerical results, 
 it has been observed that the largest eigenvalue is given by the first $k \times k$ submatrix $ A^{(0)} _{\{ 0,1,2, \cdots ,  k-1 \} }$, namely, $\max_J \max_{|\vec n|=k} \|  A^{(J)} _{\vec n } \|  =  \|  A^{(0)} _{\{ 0,1,2, \cdots ,  k-1 \} } \|  $.
Moreover, we can reproduce   the expressions of the  classical limit \cite{Namiki07,Namiki11} and the quantum limit \cite{Namiki11R}   
 from the subspace of $J=0$ %
 for $k=1$ and $k = \infty$.
 To be concrete, it holds that  
\begin{eqnarray}
 F^{(1)}(1/N,   \tau /N )&=& \frac{N+\tau }{1+N+\tau} \| A^{(0)}_{\{0\}} \|= \frac{N+\tau }{1+N+\tau}, \nonumber \\ 
     F^{( \infty)} (1/N,   \tau /N) & \leq& \frac{N+\tau }{1+N+\tau} \| A^{(0)} \|  \nonumber \\ &=& \frac{  (N+\tau +1)-|N+\tau -1|}{2}. 
\end{eqnarray}
 We thus make a conjecture  that the general limit is given by a significantly simple form: 
\begin{eqnarray}
F^{(k)}(1/N,   \tau /N) \le \frac{N+\tau }{1+N+\tau} \|  A^{(0)} _{\{ 0,1,2, \cdots ,  k-1 \} } \|. \label{conjecture}
\end{eqnarray}

Regarding the tightness of this conjectured bound, we present  a $k$-PEB channel which saturates the inequality of 
Eq.~(\ref{conjecture}) when $\tau \to 0$. Let us define a $k$-PEB channel
$\mathcal E ^{(k)} (\rho ) = \int d^2 \alpha  K_\alpha^{(k)} \rho {( K_\alpha^{(k)})}^\dagger $ with Kraus operators of rank $k$ or less-than $k$
\begin{align}
K_\alpha^{(k)} :=  \frac{1}{\sqrt \pi}  D \left(  \frac{\sqrt{\eta} \alpha}{1+\lambda }  \right) \left(\sum_{n=0}^{k-1} a_n^{(k)} \ketbra{n}{n}  \right) D^\dagger ( \alpha) . 
\end{align}
It fulfills $\int d^2 \alpha  {(K_\alpha^{(k)})}^\dagger   K_\alpha^{(k)} 
 = \openone$ 
 by  imposing the condition 
  $\sum_{n=0}^{k-1} a_n^2 = 1$ and  
 gives a simple form of the fidelity  \begin{eqnarray}
\lim_{\lambda \to 0 } F_{ \eta, \lambda } (\mathcal E^{(k)}) 
\nonumber   &= & \sum_{n,m=0}^{k-1} \frac{ a_n a_m \sqrt{\eta}^{n+m}}{n! m!} \left(\frac{-\partial}{\partial \eta}\right) ^{n+m }\!\! \frac{1}{1+ \eta} \\
&=&  \frac{1}{1+ \eta} \sum_{n,m=0}^{k-1}  a_n \gamma_{n,m}^{(0)} a_m   =: f ^{(k)},
\end{eqnarray}
where $\gamma_{n,m}^{(0)}$ is given by Eq.~\eqref{gammanm} with $x= \sqrt{\eta }/(1+\eta)$.
This implies $\max_{\{ a_n \} } f ^{(k)}  
= (1-\eta)^{-1}
 \|A_{\{0, 1, \cdots , k-1  \}} ^{(0)}  \| $,
  and  $\mathcal E^{(k)}$  achieves the conjectured bound of Eq.~(\ref{conjecture}) for $\tau =0$. 
It was shown that  $\mathcal E^{(1)} $
achieves the classical limit in Ref.~
\cite{Namiki07}.  For $k=2$ and $k=3$, we have observed numerically that $\mathcal E^{(k)} $ could not achieve the limit when $\tau >0$. Interestingly, we can generally show  that the conjectured fidelity bound in Eq.~\eqref{conjecture} is achievable by a probabilistic quantum gate of Schmidt-class $k$ for every $k$ (See  Sec~\ref{APA2}).

\section{Extension for general quantum operations}

Our benchmarks can be  extended for general quantum operations, namely, trace-non-increasing class of CP maps (See  Ref.~\cite{Namiki1503} for a general framework).  
  In Sec.~\ref{APA1} we show that the bound $U_k$ of Eq.~\eqref{Obound} holds for CP maps of Schmidt-class $k$ with a modified form of the fidelity. Notably, the bound $U_k$ is tight when general quantum operations are taken into account.  
An interesting example of  trace-decreasing CP maps for CV states is 
 the so-called  noiseless linear amplifier  or probabilistic amplifiers 
\cite{Ralph09,Chrzanowski2014,Xiang2010a,Neergaard-Nielsen2013,Namiki1502,Chir13}.  In Sec.~\ref{APA2}, we  prove that  such a stochastic quantum channel achieves  the conjectured bound of Eq.~\eqref{conjecture}.

\label{APA}
\subsection{Fidelity bounds for CP maps}\label{APA1}
Suppose that $\mathcal E$ is a quantum operation, namely, a trace-non-increasing CP map. We may modify the definition of the fidelity in  Eq.~\eqref{eq1}  as \cite{Chir13} 
\begin{align}
&F_{\eta, \lambda} (\mathcal E)  :=  P_s^{-1}  
 \int p_\lambda( \alpha )\bra{\sqrt \eta \alpha }\mathcal E \left(| \alpha \rangle \langle \alpha| \right) \ket{\sqrt\eta \alpha} d^2 \alpha  \label{eq1d},& \end{align}
where  $p_\lambda( \alpha ) =  \frac{\lambda }{\pi} \exp (- \lambda |\alpha |^2 )$ with  $\lambda >0 $, and  $ P_s := { \tr\int p_\lambda (\alpha)     \mathcal E (\ketbra{\alpha}{\alpha})  d^2 \alpha}$ is  the probability that  $\mathcal E$ gives an output state  for the ensemble $\{p_\lambda (\alpha), \ketbra{\alpha}{\alpha} \}$. 
Note that Eq.~\eqref{eq1d} reduces to  Eq.~\eqref{eq1} for the trace-preserving case, i.e.,  for  quantum channels. In fact, $\tr  [   \mathcal E (\ketbra{\alpha}{\alpha})] =1 $ for all $\alpha \in \mathbb{C}$ implies $P_s =1$.

Similarly to Eq.~\eqref{defk}, let us define the Schmidt-class-$k$ fidelity limit with the renormalized fidelity in  Eq.~\eqref{eq1d} as 
 \begin{eqnarray}
F^{(k)}(\eta, \lambda) 
 &:=& \max_{\mathcal E \in \mathcal O_k } F_{\eta, \lambda} (\mathcal  E)  , \label{defk-} 
\end{eqnarray} where $\mathcal O_k$ denotes  the set of $k$-PEB maps described by $\mathcal E (\rho ) = \sum_i K_i \rho K_i^\dagger $. Here, operators $\{ K_i \} $ have rank $k$ or less than $k$ (We do not impose trace-preserving condition $\sum _iK_i^\dagger K_i = \openone $).

In order to show that the same fidelity bound $U_k$ in Eq.~\eqref{Obound}
holds for quantum operations \cite{Namiki1503}, 
the key is to employ the normalized state: 
\begin{eqnarray}
\rho_{\mathcal E} :&=& \frac{ \mathcal E \otimes I \left( \ket{\psi_\xi}\bra{\psi_\xi} \right )}{\tr\left[  \mathcal E \otimes I \left( \ket{\psi_\xi}\bra{\psi_\xi} \right )\right]}. \label{normrho} 
\end{eqnarray} By using this formula, instead of  Eq.~\eqref{cms},
  the procedure in Sec.~\ref{bounding} leads to  the fidelity bound for general CP maps:  
\begin{align}
F^{(k)}(\eta, \lambda ) \le  U_k(\eta, \lambda )  = \frac{1+\lambda }{1+\eta+\lambda}  \max_J \max_{|\vec n|=k} \|  A^{(J)} _{\vec n } \| ,   \label{18ap}
\end{align} where $\| \cdot \|$ denotes the  maximum eigenvalue and  $ A^{(J)} _{\vec n } $ is defined through Eqs.~\eqref{gammanm}, \eqref{Amat}, and \eqref{defA} with 
\begin{align}
x = \frac{\sqrt{\eta (1+ \lambda)}}{1+ \eta +\lambda}, \ \kappa = \sqrt\frac{1+\lambda}{\eta }. \label{xxx} \end{align}
Note that Eq.~\eqref{18ap} is tight, namely, it holds that
\begin{align}
F^{(k)}(\eta, \lambda )=  U_k(\eta, \lambda ). 
\end{align}  This can be confirmed from the fact that $ U_k$ is a solution of the Schmidt-number-eigenvalue problem \cite{Sperl11} together with the property of $k$-PEB maps that   $\rho_{\mathcal E}$ of Eq.~\eqref{normrho} can be any pure state of Schmidt-number $k$. 
For quantum channels (trace-preserving CP maps), it remains open how to find tight limit except for the classical limit $k=1$ \cite{Namiki07} and quantum limit $k= \infty$ \cite{Namiki11R,Chir13}.

By comparing an experimentally observed fidelity and our upper bound of the $k$-th fidelity limit $F^{(k)}$, one can verify a genuine multi-dimensional coherence for general quantum operations as well as for quantum channels.  To be concrete, we can eliminate the possibility that the physical process $\mathcal E$ is described as a $k$-PEB map if it holds that $F_{\eta, \lambda} (\mathcal E) > U_k (\eta,\lambda )$.  This establishes  an infinite sequence of quantitative quantum benchmarks for general single-mode physical processes    with respect to the Schmidt number, $k$.  
The fidelity steps $U_k (\eta,\lambda )$ would give distinctive milestones to  assess the closeness between an experimental amplifier and an ideal quantum limited amplification process \cite{Chir13,Namiki1502} by simultaneously observing the Schmidt number and the fidelity.

\subsection{Proof of attainability of the conjectured bound} \label{APA2}
In Sec.~\ref{subseccunjecture}, we have conjectured that the simple form in Eq.~\eqref{conjecture} gives a tighter bound. 
Here, we will show that  a probabilistic quantum channel of Schmidt-class $k$ achieves the conjectured bound: 
\begin{eqnarray}
F^{(k)}(\eta, \lambda ) \le \frac{1+\lambda }{1+\eta+\lambda}  \|  A^{(0)} _{\{ 0,1,2, \cdots ,  k-1 \} } \|.    \label{ConjectureBound}
\end{eqnarray}

\textit{Proof.---}
Let us consider the following filtering operator
\begin{align}
Q_k =  \sqrt{\mathcal N }  \sum_{n=0}^{k-1} a_n g^n \ketbra{n}{n} \label{modefil}
\end{align} where $(\mathcal N, g $)  is a pair of positive constants and we assume $\sum_{n=0}^{k-1}| a_n| ^2 =1 $. We can readily calculate its action onto coherent states as 
\begin{align}
 Q_k \ket{\alpha}  
  = \sqrt{\mathcal N}  e^{- |\alpha |^2/2}  \sum_{n=0}^{k-1}  \frac{a_n  ( g  \alpha   ) ^n }{\sqrt{ n!} }  \ket{n}.  \label{QK} 
\end{align}

Evidently, $Q_k$ is rank $k$ or less than $k$. This implies that the probabilistic quantum gate $\mathcal E (\rho)=  Q_k \rho Q_k ^\dagger  $ belongs to Schmidt-class $k$. From these expressions we have   
\begin{align}
& \int p_\lambda( \alpha )\bra{\sqrt \eta \alpha }\mathcal E \left(| \alpha \rangle \langle \alpha| \right) \ket{\sqrt\eta \alpha} d^2 \alpha \nonumber \\
=& \mathcal N  \lambda  \sum_{n,m =0}^{k -1}\frac{(n+m)!}{n! m!} \frac{(\sqrt \eta g )^{m+m } a_n a_m ^*}{(1+ \eta + \lambda)^{n+m+1}} \nonumber \\ 
 = &\frac{  \mathcal N  \lambda}{ 1+ \eta + \lambda } \sum_{n,m =0}^{k -1}  \gamma_{n,m} ^{(0)} a_n a_m ^* , \label{eqa9}
\end{align}
where we use $\int p_\lambda(\alpha) e^{-(1+ \eta )|\alpha |^2}  |\alpha |^{2k} d^2\alpha =  \lambda {k!}/{(1+\eta +  \lambda )^{k+1}}$ for calculating  the integration. Moreover, the final expression is obtained by substituting $g = \sqrt {1+ \lambda }$ and using  the definition of $\gamma_{n,m}^{(0)} $ in  Eq.~\eqref{gammanm} where the parameter $x$ is given by Eq.~\eqref{xxx}. 
Note that, from the definition of the submatrix $A_{\vec n}^{(J)}$ given through  Eqs.~\eqref{gammanm},~\eqref{Amat},~and~\eqref{defA},  we can write 
\begin{align}
\max 
 \left(   \sum_{n,m =0}^{k -1} { \gamma_{n,m} ^{(0)} a_n a_m ^*} \right) = \| A_{ \{ 0,1,2, \cdots, k-1 \} }^{(0)} \| , \label{maxa}
\end{align}
where the maximum is taken over $\sum_{n=0}^{k-1}| a_n| ^2 =1 $.

On the other hand, the relation in Eq.~\eqref{QK} and the condition $g =\sqrt{1+ \lambda} $ yield the following expression: 
\begin{align}
P_s & = { \tr\int p_\lambda (\alpha)     \mathcal E (\ketbra{\alpha}{\alpha})  d^2 \alpha} \nonumber \\  
& =\mathcal N \lambda  \sum_{n=0}^{k -1}  \frac{ |a_n | ^2 g ^{2n }}{(1+\lambda )^{n+1} } 
=\frac{\mathcal N \lambda}{1+ \lambda}  \sum_{n=0}^{k -1}    |a_n | ^2 =\frac{\mathcal N \lambda}{1+ \lambda}  .  \label{psfinal}
\end{align}
 From Eqs.~\eqref{eq1d},~\eqref{eqa9},~and~\eqref{psfinal} we obtain
 \begin{align}
F_{\eta, \lambda} (\mathcal E)  = \frac{1 +  \lambda}{ 1+ \eta + \lambda } \sum_{n,m =0}^{k-1}  \gamma_{n,m} ^{(0)} a_n a_m ^*.  
\end{align} Finally, optimizing the coefficient $\{a_n\}$ of the filter $Q_k$ as in Eq.~\eqref{maxa} 
 we can conclude  that the right-hand side of Eq.~\eqref{ConjectureBound} [Eq.~\eqref{conjecture}] is achievable by a probabilistic quantum gate of Schmidt-class $k$.  
 \hfill$\blacksquare$

Note that the normalization factor $\mathcal N$ of $Q_k$ in  Eq.~\eqref{modefil} can be positive as long as  $k$ is finite.  This fact confirms the attainability with a finite success probability $P_s >0$. 
In the limit of  $k \to \infty$, however, $\mathcal N $ could be zero so as to fulfill the physical condition $Q_k^\dagger Q_k \le \openone$ (See Refs.~\cite{Chir13,Namiki1502}).

\section{Schmidt-class-$k$ limitation on quantum noise of canonical variables}
In this section, we present a Schmidt-class-$k$ limit on an average of Bayesian mean-square deviations for canonical variables.  We introduce a basic relation between the fidelity and quantum noise in Sec.~\ref{QNandF}. Resultant Schmidt-number benchmarks are given in  Sec.~\ref{QNBen}.

\subsection{Canonical quantum noise and fidelity}\label{QNandF}
Let $\hat x$ and $\hat p$ be canonical quadrature variables with the canonical commutation relation $[\hat x,\hat p] =i$.
The field operator $\hat a $ is given as  $\hat a = (\hat x + i\hat p)/\sqrt 2$,  and  satisfies the bosonic commutation relation, $[\hat a, \hat a ^\dagger ]=1$.  For notation convention we write 
the mean quadratures for coherent states as   \begin{align}
x_\alpha  :=  \bra{\alpha}\hat x  \ket{\alpha}= \frac{\alpha + \alpha ^*}{\sqrt 2 }, \  p_\alpha  :=   \bra{\alpha}\hat p \ket{\alpha}= \frac{\alpha - \alpha ^*}{\sqrt 2 i }.  \label{Eq2}\end{align} 

Let $\mathcal E$ be a quantum operation. We define the mean-square deviations for canonical quadratures \cite{Namiki07,Namiki-Azuma13x,Namiki1502} as 
\begin{align} \bar V_z   :=  P_s^{-1} \tr \left[  \int  p_\lambda (\alpha )   (\hat 
z-  \sqrt \eta  z_\alpha )^2   \mathcal E ( \ketbra{\alpha}{\alpha} ) d^2 \alpha \right],  & \label{MSD}    \end{align} where $z \in \{x,p\}$, 
  $p_\lambda( \alpha ) :=  \frac{\lambda }{\pi} \exp (- \lambda |\alpha |^2 )$, and $ P_s := { \tr\int p_\lambda (\alpha)     \mathcal E (\ketbra{\alpha}{\alpha})  d^2 \alpha}$.  With the help of   the property of the displacement operator   $ D( \alpha ) \hat a  D^\dagger  ( \alpha ) =\hat a-   \alpha$ and  the cyclic property of the  trace,  we can write  \begin{align} \bar V_z  = & P_s^{-1} \tr \left[  \int  p_\lambda (\alpha )      D(\sqrt \eta \alpha ) \hat z ^2  D^\dagger  (\sqrt \eta \alpha )  \mathcal E ( \ketbra{\alpha}{\alpha} ) d^2 \alpha \right]\nonumber \\ 
  =& P_s^{-1} \tr [ \hat z^2 \sigma ]  ,  \label{MSDR}    \end{align}
where we  defined 
\begin{align}
\sigma: = &   \int  p_\lambda (\alpha )  D^\dagger  (\sqrt \eta \alpha )    \mathcal E ( \ketbra{\alpha}{\alpha} )  D(\sqrt \eta \alpha )   d^2 \alpha  . \label{sigma} 
\end{align}  
Note that we can readily confirm  the following relations: 
\begin{align}
\tr [\sigma] = & \tr \int  p_\lambda (\alpha )   \mathcal E ( \ketbra{\alpha}{\alpha} )      d^2 \alpha  = P_s,  \nonumber \\ 
\bra{0}\sigma \ket{0}  = &  \int  p_\lambda (\alpha ) \bra{\sqrt \eta \alpha }      \mathcal E ( \ketbra{\alpha}{\alpha} )  \ket{\sqrt \eta \alpha} d^2  \alpha 
. \label{sigma1} 
\end{align}

From Eq.~\eqref{MSDR} and the well-known expression for the harmonic oscillator, $\hat x^2 + \hat p^2=  2  \hat  a ^\dagger \hat a +1 $, 
 the sum of the mean-square deviations can be expressed as  
\begin{align}
\bar V_x+\bar V_p    = &     \frac{1}{P_s} (2  \tr  [\hat  a ^\dagger \hat a  \sigma] +  \tr  [  \sigma]) \label{vsum} .  
\end{align}
On the other hand, we can show  the following inequality for any positive semidefinite operator $\rho$:  
\begin{align}  
 \tr [\hat a^\dagger\hat a \rho  ] = &\tr \left(  \sum_{n=1}^\infty n \ketbra{n}{n} \rho \right )  \nonumber \\  \ge & \tr \left(  \sum_{n=1}^\infty  \ketbra{n}{n} \rho \right )= 
  \tr [ \rho ] - \bra{0}\rho \ket{0} .  \label{ThisRelation}\end{align}
Concatenating  Eqs.~\eqref{sigma1},~\eqref{vsum},~and~\eqref{ThisRelation} 
with $\rho = \sigma$, we obtain the relation between the sum  quantum noise and the average fidelity \cite{Namiki07}
\begin{align}
\bar V_x+\bar V_p -1  
\ge&   \frac{2}{P_s}  (\tr [\sigma ] - \bra{0} \sigma \ket{0}) = 
2\left(  1- F_{\eta,\lambda }   \right),   \label{vsumeq}
\end{align}
 where $F_{\eta,\lambda } $ is defined in Eq.~\eqref{eq1d}.  
From Eq.~\eqref{vsumeq}, we can see that a smaller value of  quantum noise ensures  a higher fidelity.  To be concrete, Eq.~\eqref{vsumeq} implies that  the fidelity is bounded from below by using  the mean-square deviations 
\begin{align}
 F_{\eta,\lambda } \ge& \frac{3}{2}  - \frac{ \bar V_x+\bar V_p}{2} . 
\end{align} 
In particular,  we can observe that $F=1$  if  $V_x +V_p =1$.

\subsection{Schmidt-number benchmarks via quantum noise} \label{QNBen}
Now, we can find a  lower bound of the Schmidt number by using the  sum of the mean-square deviations $V_x $  and $V_p$.   
We can show that  $F_{\eta, \lambda} (\mathcal E) >  F^{(k)}(\eta, \lambda)$ holds  
 if  the following condition is satisfied
\begin{align} 
 \bar V_x+\bar V_p-1  < 2(1-F^{(k)}(\eta, \lambda) ).  \label{quadrabound} 
\end{align}  

\textit{Proof.---} Suppose Eq.~\eqref{quadrabound} holds. 
From  Eqs.~\eqref{vsumeq}~and~\eqref{quadrabound}, we have
\begin{align} 
2\left(  1- F_{\eta,\lambda }   \right) \le  \bar V_x+\bar V_p-1  < 2(1-F^{(k)}(\eta, \lambda) ). \end{align}
Comparing the left-end and right-end expressions, we obtain $F_{\eta, \lambda} (\mathcal E)  >  F^{(k)}(\eta, \lambda)$.  Hence, Eq.~\eqref{quadrabound}  is a sufficient
condition that $\mathcal E$ outperforms any $k$-PEB maps. 
 \hfill$\blacksquare$

For a practical use, one can replace the term $F^{(k)}$ in Eq.~\eqref{quadrabound} with the upper bound $U_k $ given in Eq.~\eqref{Obound}. We thus have the following quantum benchmark: 
\begin{align} 
 \bar V_x+\bar V_p-1  < 2(1-U_k (\eta, \lambda) ).  \label{quadrabound1} 
\end{align}  
This condition 
can be readily tested by plugging-in an experimentally observed  value of $\bar V_x+ \bar V_p$.  
Hence,  one can verify that the Schmidt number of the process $\mathcal E$ is at least $k+1$ if the inequality of Eq.~\eqref{quadrabound1} is fulfilled. 

Note that the condition of  Eq.~\eqref{quadrabound} is not tight when $k=1$ (See, Corollary~1 of  \cite{Namiki-Azuma13x}), and unlikely to be tight for other choice of  $k \ge  2$.  
How to find a better link between  the Schmidt class and the quadrature deviations for an improvement of our approach should be addressed elsewhere.
  

\section{Conclusion and remarks}
In conclusion, we have presented Schmidt-number benchmarks for CV quantum devices using the average fidelity for Gaussian distributed coherent states.  Our benchmarks give everlasting fidelity steps towards higher dimensional quantum-gate coherence, and successfully generalize the known classical and quantum limits by recasting them into the two endpoints of these steps.  Our result refines the meaning of  ``high fidelity'' for CV quantum gates, and 
the numerically determined fidelity steps would be useful to demonstrate genuinely higher dimensional coherence for 
  experimental implementations. %
  It is  fundamentally  important to show stronger evidence that  CV systems have a potential superiority in dealing with  higher dimensional quantum signals. In this respect, distinctive experimental progress could be regularly quantified and recorded based on the Schmidt class determined by  the fidelity steps. 
 We have also conjectured a simple formula for the fidelity bound. This bound is achievable by a probabilistic quantum gate of the corresponding  Schmidt class.  Furthermore, we have  presented a lower bound of an average quantum noise to outperform $k$-PEB processes. This bound is directly related to homodyne measurements and could  provide wider options for an experimental verification of higher dimensional coherences.
 
 Although our results are readily available  as a type of entanglement verification tools for experiments, there are several open possibilities to improve the fidelity bound and the bound for the canonical quantum noise. We remark the following three aspects for an outlook. 

 (i)~Our  fidelity bound $U_k$ is tight for quantum operations, yet we have not identified what operation can  achieve this  bound (In Sec.~\ref{APA2} we have provided a concrete form of a  probabilistic gate that achieves the conjectured bound. If the conjectured bound is proven equivalent to $U_k$,  we can immediately settle this problem).
 
 (ii)~How to improve the fidelity bound $U_k$ for the case of quantum channels and how to identify the optimal $k$-PEB channel which  maximizes the fidelity, for $k\in [2, \infty)$ remain open. In this regard, it has been known \cite{Chir13} that there is a gap between the quantum limit fidelities ($k \to \infty$) for probabilistic gates and deterministic gates, whereas there is no gap for the classical fidelity limits ($k=1$). An existence of the gap is crucial  to demonstrate an advantage of probabilistic gates \cite{Namiki1502}. 
 
(iii)~Aside from the fidelity-based approach, 
 exploring a feasible method based on the statistical moments of canonical variables would be important. As well as improving our bound for the sum of the mean-square deviations in Sec.~\ref{QNBen}, an interesting problem is to determine  the trade-off relation between the mean-square deviations under the constraint of the Schmidt class. 
Hopefully, we could prove a general sequence of uncertainty relations for $V_x $ and $V_p$, which reproduces the uncertainty relation over Entanglement-breaking maps for $k=1$ \cite{Namiki-Azuma13x} and approaches the amplification uncertainty relation \cite{Namiki1502} in the limit $k \to \infty$.



\acknowledgments
RN was supported by the DARPA Quiness program under prime Contract No. W31P4Q-12-1-0017, NSERC, and Industry Canada. This work was partly supported by GCOE Program ``The Next Generation of Physics, Spun from Universality and Emergence'' from MEXT of Japan. 


\appendix



\section{Rigorous result for the maximization in Eq.~\eqref{Obound} for $J=0$}\label{APB}
As a first step to estimate the maximization in Eq.~\eqref{Obound},  we consider the case of $J=0$. 
In the case of $J=0$ 
we can show that 
the exact maximum for $k=1$, 2, and 3   
is given by 
 \begin{eqnarray} \max_{|\vec n|=k } \|  A^{(0)} _{\vec n } \| &=&  \|  A^{(0)} _{\{0,1,2, \cdots, k-1 \} } \|. 
 \label{kEq3} \end{eqnarray}
 In order to verify this relation, we use the following two properties for $\gamma$ defined through  Eqs.~\eqref{gammanm}~and~\eqref{definitionx}: 
\\(i) $\gamma_{n+1,n-l+1}^{(0)}-\gamma_{n,n-l}^{(0)}\le 0$ holds for  $n \ge \frac{1}{2} (l+2)(l-l)$.  
 \\(ii) $\gamma_{n+1,m}^{(0)}-\gamma_{n,m}^{(0)} \le 0$ holds for $m \le n-1 $.

First, Property (i) with  $l=0$ implies  that the diagonal elements are in decreasing order, namely,  $\gamma_{0,0}^{(0)} \ge \gamma_{1,1}^{(0)} \ge \gamma_{2,2}^{(0)} \ge \cdots $. This proves Eq.~(\ref{kEq3}) for $k=1$.  Note that Property (i) with  $l=1$ implies  that the first off-diagonal elements are in  decreasing order, namely, it holds that $\gamma_{0,1}^{(0)} \ge \gamma_{1,2}^{(0)} \ge \gamma_{2,3}^{(0)} \ge \cdots $. Similarly,  Property (i) with  $l=2$ implies  that the second off-diagonal elements are in decreasing order, namely,  it holds that  $\gamma_{0,2}^{(0)} \ge \gamma_{1,3}^{(0)} \ge \gamma_{2,4}^{(0)} \ge \cdots $. 

\begin{figure}[hptb]
\includegraphics[width= \linewidth]{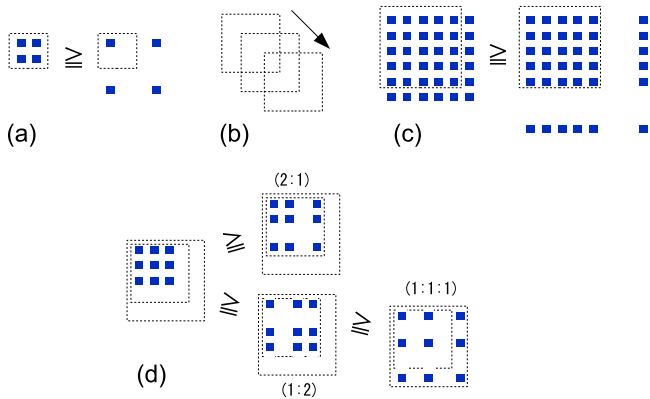} 
  \caption{Orders of principal submatrices.}  \label{fig:1}
\end{figure}

Next, to prove Eq.~(\ref{kEq3}) for $k=2$ we show
\begin{eqnarray}
\left(
  \begin{array}{cc}
    \gamma_{n,n}^{(0)}     &  \gamma_{n,n+1}^{(0)}   \\
    \gamma_{n+1,n}^{(0)}    &   \gamma_{n+1,n+1}^{(0)}   \\
  \end{array}
\right) - 
\left(
  \begin{array}{cc}
    \gamma_{n+1,n+1}^{(0)}     &  \gamma_{n+1,n+2}^{(0)}   \\
    \gamma_{n+2,n+1}^{(0)}    &   \gamma_{n+2,n+2}^{(0)}   \\
  \end{array}
\right)  \ge 0\nonumber , \\
\left(
  \begin{array}{cc}
    \gamma_{n,n}^{(0)}     &  \gamma_{n,n+1}^{(0)}   \\
    \gamma_{n+1,n}^{(0)}    &   \gamma_{n+1,n+1}^{(0)}   \\
  \end{array}
\right) - 
\left(
  \begin{array}{cc}
    \gamma_{n,n }^{(0)}     &  \gamma_{n ,n+n'}^{(0)}   \\
    \gamma_{n+n',n+n'}^{(0)}    &   \gamma_{n+n',n+n'}^{(0)}   \\
  \end{array}
\right)  \ge 0 , \nonumber \\ \label{posiposp}\end{eqnarray}
where each inequality for matrices indicates all elements are non-negative.
The first inequality suggests the decreasing order on shift in the diagonal direction  associated with the schematics of Fig.~\ref{fig:1}(b); The second inequality suggests the decreasing order on spread in the vertical-and-horizontal direction associated with the schematics of Fig.~\ref{fig:1}(a).  The first inequality of Eqs.~(\ref{posiposp}) is proven from the decreasing order on the diagonal elements and the first off-diagonal elements. The second inequality  of Eqs.~(\ref{posiposp}) is proven by  using the decreasing order on the diagonal elements and Property (ii). From the inequalities in Eqs.~(\ref{posiposp}) we have $ \|  A^{(0)} _{\{n,n+1\} } \| \ge  \|  A^{(0)} _{\{n+1,n+2\} } \|$ and $ \|  A^{(0)} _{\{n,n+1\} } \| \ge  \|  A^{(0)} _{\{n,n+n'\} } \|$  since $\| a \| \le \| b \|  $ holds for non-negative matrices $a$ and $b$ with $0 \le a\le b $ (See, Corollary 8.1.19. of \cite{MatAnn}). Using these two relations recursively we can conclude Eq.~(\ref{kEq3}) for $k=2$. 
Note that,  from the order of the diagonal elements and  Property (ii),  we can generally obtain such a matrix inequality when the position of the final row and column is shifted as in Fig.~\ref{fig:1}c.       

Finally, similar to this proof, we  proceed to the proof of $k=3$
by comparing the corresponding submatrix elements associated with $  \|  A^{(0)} _{\{n,n+1,n+2\} } \| \ge  \|  A^{(0)} _{\{n+1,n+2,n+3\} } \|$ for the diagonal direction shift and $  \|  A^{(0)} _{\{n,n+1,n+2\} } \| \ge  \|  A^{(0)} _{\{n,n+n',n+n''\} } \|$ for the spreading shift. For the diagonal shift of the $3 \times 3$ matrix, the matrix inequality can be confirmed by  the decreasing order on the diagonal elements, the first off-diagonal elements, and the second off-diagonal elements, coming from Property (i) with $l=0$, $1$, and $2$. For the spreading of 
the $3\times 3$ matrix, we have three possibilities to divide the elements (2:1), (2:1), and (1:1:1) as in Fig.~\ref{fig:1}(d). For the  case of (2:1), the inequality can be proven by  Property (ii) and the decreasing order on the diagonal elements.  For the  case of (1:2), the inequality can be proven by Property (ii)  and the decreasing order on the diagonal shift of $2\times  2$ matrix.  Then, the first inequality of Eq.~(\ref{posiposp}) on $2 \times 2$ matrix and the decreasing order on the diagonal elements again enable us to show the decreasing order on the spreading shift from (1:2) to (1:1:1).  
Therefore, we can conclude that the relation Eq.~\eqref{kEq3} holds for $k \in \{1,2. 3\}$.  

For $k=4$, we can show the inequality  for the spreading shift 
 $  \|  A^{(0)} _{\{n,n+1,n+2,n+3\} } \| \ge  \|  A^{(0)} _{\{n,n+n',n+n'',n+n'''\} } \|$ by using Property (ii) and the results of the $2\times 2$ and $3 \times 3$ matrices above. Similarly, from Property (i) and the results of $k \le 3$ above we have $  \|  A^{(0)} _{\{n,n+1,n+2,n+3\} } \| \ge  \|  A^{(0)} _{\{n+1,n+2,n+3,n+4\} } \|$ when $n \ge 2$. However, the matrix inequality for the diagonal shift could not hold for the first two submatrices, $A_{\{0,1,2,3\}}^{(0)}$ and $A_{\{1,2,3,4\}}^{(0)}$. Therefore, the maximum is obtained by comparing the first three cases of the matrices, i.e. $\max_{n\in \{ 0, 1,2 \}} \| A_{\{n,n+1,n+2,n+3\}}^{(0)}\|$. In this manner, 
  we can eventually determine the maximum by comparing the maximum eigenvalues  of a relatively small number of submatrices for a couples of small $k$. We present a general systematic approach to determine the maximum of Eq.~\eqref{Obound} in the following  section. 

\section{General recipe to determine the  maximum in Eq.~\eqref{Obound}}\label{APC}
In the previous section, we use the following two properties to make (matrix) inequalities on submatrices of $A^{(0)}$ defined through Eqs.~(\ref{gammanm})~and~(\ref{definitionx}):
\\ (i) $\gamma_{n+1,n-l+1}^{(0)}-\gamma_{n,n-l}^{(0)}\le 0$ holds for  $n \ge \frac{1}{2} (l+2)(l-l)$.  
 \\(ii) $\gamma_{n+1,m}^{(0)}-\gamma_{n,m}^{(0)} \le 0$ holds for $m \le n-1 $. 

In this section, we develop  this method and present a systematic approach to determine the  maximum in Eq.~\eqref{Obound}. 
 An essential fact to generate matrix inequalities is that $\| a \| \le \| b \|  $ holds for non-negative matrices $a$ and $b$ with $0 \le a\le b $ (See, Corollary~8.1.19. of \cite{MatAnn}). 

Let us note general properties of $A^{(J)}$. (a)~$A^{(J)}$ is a non-negative matrix and symmetric, i.e., for any $n,m$, $\gamma _{n,m}^{(J)} \ge 0$ and  $\gamma _{n,m}^{(J)}=\gamma _{m,n}^{(J)}$. 
 (b)~If $n \ge \frac{1}{2}(|J|-2)(|J|+1)$, 
 we have $\gamma_{n+1,n+1}^{(J)}-\gamma_{n,n}^{(J)}\le 0$.   
 (c)~The sequence of the diagonal elements $\{\gamma_{n,n}^{(J)}\}_n $ is at most single peaked and the largest element is located around $n \sim \frac{1}{2} (|J| ^2-|J|-2)$. From these properties and the fact that eigenvalues of a positive semidefinite matrix are upper bounded by its trace, we can neglect the contribution from  sufficiently large  $n$ when we  determine the  maximum in Eq.~\eqref{Obound}, numerically. 

\subsection{Derivation of the properties (i) and (ii)} \label{APC1}
From Eqs.~\eqref{gammanm}~and~\eqref{definitionx}  we have
\begin{align}
&  \gamma_{n+1,m+1}^{(J)}-\gamma_{n,m}^{(J)} \le 
\gamma_{n,m}^{(J)} \nonumber \\
&  \times \left( \frac{( n+m+|J|+2)(n+m+|J|+1)}{\sqrt{(n+1)(m+1)(n+|J|+1)(m+|J|+1)}}\frac{1}{4}-1 \right). 
\end{align} Suppose that $m= n-l \le n $. If we set $J=0$ we obtain Property (i).  
In our approach, a key observation is that any $l$-th off-diagonal element gradually gives a decreasing sequence. We define an integer $t_l$  to utilize this fact. 
 The integer $t_l$ that fulfills $\gamma_{n+1,m+1}^{(J)}-\gamma_{n,m}^{(J)}\le 0 $ is summarized in Table \ref{table1} for $l \leq 10$. The row of $l=0$ shows  all diagonal elements are in decreasing order for $|J| \le 2$.  The row of $l=1$ shows  all first off-diagonal elements are in decreasing order for $|J| \le 1$. Note that the values in Table \ref{table1} are determined by  taking the  worst case of $x= \frac{1}{2}$  (Better bounds would be obtained 
 when a specific value of $x<\frac{1}{2} $ is given).

 Suppose that $m \le  n-1$. From Eqs.~(\ref{gammanm})~and~(\ref{definitionx}) we have
\begin{eqnarray}
\gamma_{n+1,m}^{(J)}-\gamma_{n,m}^{(J)} &\le& \gamma_{n,m}^{(J)}\left( \frac{n+m+|J|+1 }{\sqrt{(n+1)(n+|J|+1)}}\frac{1}{2}-1 \right) \nonumber \\
\nonumber \\
 &\le& \gamma_{n,m}^{(J)}\left( \frac{n+|J|/2 }{\sqrt{(n+1)(n+|J|+1)}} -1 \right). \nonumber \\
\end{eqnarray}
This implies $\gamma_{n+1,m}^{(J)}-\gamma_{n,m}^{(J)} \le 0$ for $|J|\le 4$. 
For $|J| > 4$, $\gamma_{n+1,m}^{(J)}-\gamma_{n,m}^{(J)} \le 0$  is fulfilled  when 
\begin{eqnarray}
n \ge (-4 - 4 |J| + J^2)/8 =: u^{(J)}. \label{breve}
\end{eqnarray}
As a consequence,  the case of $J=0$ gives Property (ii). Notably, the expressions derived here suggest that we can use modified versions of Properties (i) and (ii) for $J \neq 0$. Our main residual task is to make matrix inequalities systematically based on the general properties of $\{\gamma_{n,m}^{(J)}\}$. 

\begin{table}[tbp]
 \caption{The integer $t_l^{(J)}$ in which $l$-th off-diagonal elements become decreasing order for $|J| \le 4$. The superscript index ``$(J)$'' of $t_l^{(J)}$ is omitted through the text. 
   \label{table1}  } 
 \begin{center}
  \begin{tabular}{|c||c|c|c|c|c|}
    \hline
     $l$  & $J= 0$   & $|J |=1$   & $|J|=2$ & $|J|=3$  & $|J|=4$   \\
    \hline     \hline
     0  &  0  & 0   &  0 & 2& 5 \\
    \hline
     1  &  1  &  1  & 2  & 4 & 7\\
    \hline
     2  &  2  &  3  &  4 & 6 & 9\\
    \hline
     3  &  5  &  6  &  7 & 9 & 12\\
    \hline
     4  &  9  &  10  &  11 & 13 & 16 \\
    \hline
     5  &  14  &  15  &  16 & 18 & 21\\
    \hline
     6  &   20 &  21  &  22  & 24 & 27\\
    \hline
     7  &   27 &  28  &  29  & 31 & 34\\
    \hline
     8  &   35 &   36 &  37 &  39 & 42\\
    \hline
     9  &  44  &   45 &  46 & 48 & 51\\
    \hline
     10  &  54  &   55 & 56  & 58 & 61\\
    \hline
  \end{tabular}
 \end{center}
\end{table}
\subsection{Inequalities for the diagonal shift}
\begin{figure}[tb]
\includegraphics[width=0.9\linewidth]{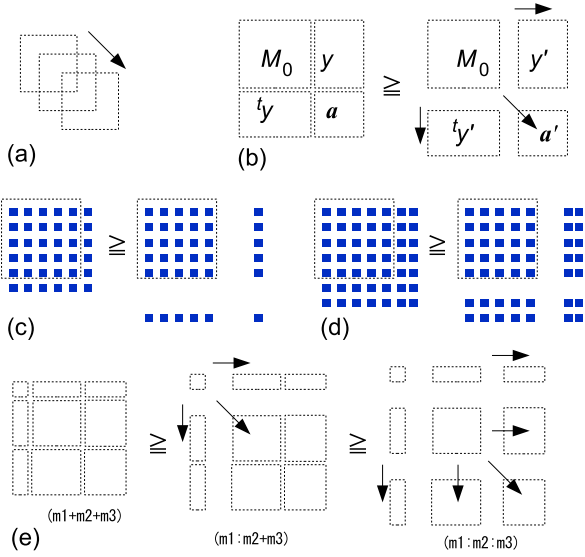}
  \caption{To determine the order of the maximum eigenvalues of the different submatrices  (a) a primitive step is to compare submatrices which can be chosen by shifts on the diagonal direction. (b) Another primitive step is  to compare submatrices which have the same elements but some of the last block have higher number indices (row and columns). (c) The case with the last one element is shifted. (d) The case with last two elements are shifted. (e) When we compare the submatrices which have many blocks we consider spreading of the largest sub-block such as $m2+m3$, firstly. Then, this submatrix, say (m1;m2+m3), can be connected with another submatrix, say (m1;m2;m3), by considering further spreading of the second largest sub-block such as $m3$.}  \label{fig:v}
\end{figure}

From Table~\ref{table1}, we can determine the index $n$ of the diagonal elements  of $ A^{(J)} $ so that the $k \times k$ principle submatrices starting from $[A^{(J)}]_{n,n}$ become decreasing order associated with the diagonal shift of Fig.~\ref{fig:v}(a). 
From the rows of  $l=0$ in Table \ref{table1}, we can confirm that the diagonal elements are in decreasing order for $|J| \le 2$. The diagonal elements of $A^{(3)}$ and $A^{(4)}$ are in decreasing order whenever $n \ge 2$ and $n \ge 5$, respectively.
From the rows of  $l=0$ and $l=1$ in Table~\ref{table1}, we can confirm that the relation on the $2 \times 2$ submatrices, 
\begin{eqnarray}
\left(
  \begin{array}{cc}
    \gamma_{n,n}      &  \gamma_{n,n+1}   \\
    \gamma_{n+1,n}     &   \gamma_{n+1,n+1}   \\
  \end{array}
\right) - 
\left(
  \begin{array}{cc}
    \gamma_{n+1,n+1}     &  \gamma_{n+1,n+2}   \\
    \gamma_{n+2,n+1}   &   \gamma_{n+2,n+2}\\
  \end{array}
\right)  \ge 0\nonumber,
\end{eqnarray}
 holds for $n \ge t_1-1   = 0 $ in the case of $|J|\in \{0, 1\}$. Moreover, this matrix inequality  holds for $n  \ge t_1-1 = 1 $ in the case of $|J|= 2 $ and for $n \ge t_1-1 =3 $ in the case of $|J|= 4 $. Note again that the inequality for matrices indicates all elements are non-negative.
 
  Similarly,  from the rows of  $l\in  \{0,1,2 \} $ in Table~\ref{table1}, we can confirm that  the relation on the $3 \times 3$ submatrices,  
\begin{eqnarray}
&&\left(
  \begin{array}{ccc}
    \gamma_{n,n}      &  \gamma_{n,n+1} &  \gamma_{n,n+2}  \\
    \gamma_{n+1,n}     &   \gamma_{n+1,n+1} &  \gamma_{n+1,n+2}  \\
    \gamma_{n+2,n}     &   \gamma_{n+2,n+1} &  \gamma_{n+2,n+2}  \\
  \end{array}
\right)  \nonumber \\ && - 
\left(
  \begin{array}{ccc}
    \gamma_{n+1,n+1}     &  \gamma_{n+1,n+2}   &  \gamma_{n+1,n+3}   \\
    \gamma_{n+2,n+1}   &   \gamma_{n+2,n+2}   &   \gamma_{n+2,n+3}\\
    \gamma_{n+3,n+1}   &   \gamma_{n+3,n+2}   &   \gamma_{n+3,n+3}\\
  \end{array}
\right)  \ge 0 \label{eqC3} ,
\end{eqnarray}
 holds for $n \ge t_{2}-2 = 0 $ in the case of $J = 0  $ and for $n \ge t_{2}-2 = 1 $ in the case of $|J|= 1$. Further, the relation of Eq.~\eqref{eqC3} holds for $n \ge 1 $ in the case of $|J|= 2 $,  $n \ge 4-2 =2 $ in the case of $|J|= 3 $, and $n \ge  7-2 = 5$ in the case of $|J| =4$. 
In this manner, we can show that the inequality for the $k \times k$ submatrices,
\begin{eqnarray}
&&\left(
  \begin{array}{ccc}
    \gamma_{n,n}      &  \cdots  &  \gamma_{n,n+k-1}  \\
    \vdots      &   \ddots &  \vdots \\
    \gamma_{n+k-1,n}     &    \cdots  &  \gamma_{n+k-1,n+k-1}  \\
  \end{array}
\right)  \nonumber \\ && - 
\left(
  \begin{array}{ccc}
    \gamma_{n+1,n+1}     &  \cdots &  \gamma_{n+1,n+k}   \\
    \vdots   &   \ddots   &   \vdots\\
    \gamma_{n+k,n+1}   &   \cdots   &   \gamma_{n+k,n+k}\\
  \end{array}
\right)  \ge 0 , 
\end{eqnarray}
 holds for $n \ge t_{l=k-1} - k+1 $. 

\subsection{Inequalities for the spreading shift}
Let us consider the following inequality for the spreading shift depicted in Fig.~\ref{fig:v}(b) \begin{eqnarray}
B-B'&:= &\left(
  \begin{array}{ccc}
    M_0     & y  \\
   y^t     &    a   \\
  \end{array}
\right)  - 
\left(
  \begin{array}{ccc}
    M_0     & y'  \\
   {y'}^t     &    a '  \\
  \end{array}
\right)  \ge 0, \label{30}
\end{eqnarray} where $M_0$, $a$, and $a'$ are square matrices. 
Suppose  that the matrices in  Eq.~(\ref{30}) are submatrices of $A^{(J)}$ of Eq.~(\ref{defA}) and that $J=0$ so that Properties (i)~and~(ii) are fulfilled. From the decreasing order on the diagonal elements and Property (ii), we can show that  the relation of Eq.~(\ref{30}) holds when the final row and column are shifted as in Fig.~\ref{fig:v} (c), in which  $a$ and $a'$ are diagonal elements, and $y$ and  $y'$ are single column vectors. From the decreasing order on the diagonal and first-off diagonal elements together with  Property (ii), we can show that the relation of Eq.~(\ref{30}) holds when the final two  rows and two columns are shifted as in Fig.~\ref{fig:v}(d), (here, $a$ and $a'$ are $2 \times 2$ matrices). Similarly, we can generate the matrix inequalities  in the form of Eq.~(\ref{30}) by using Properties (i) and (ii) for any size of $a$ whenever the diagonal shift ($a \to a'$) is in decreasing order. 

By further spreading the last lows and columns associated with the position of the square matrix $a$, we can generate inequalities with more separations as in Fig.~\ref{fig:v}(e). To make three separation $(m1:m2:m3)$, we consider the diagonal shift of the column-length $m2+m3$ square matrix, firstly, and then we spread the last square matrix of  the column-length  $m3$. We can reach any given separation by repeating these process recursively.

From Properties (i) and (ii), we can see that, for sufficiently  larger $n$, the inequalities for  the diagonal shift and spreading shift always hold. This is also the case for general $J \neq 0$ since similar properties hold with a bit complicated conditions such as Eq.~\eqref{breve} and Table \ref{table1} (See  the discussion in Appendix~\ref{APC1}). Hence, the set of submatrices we need to compare the maximum eigenvalues is a finite set of smaller-$n$-index submatrices that could not be connected by the matrix inequalities obtained  by these properties. On this basis, the search of the submatrices that have larger maximum eigenvalues can be carried out by a relatively small number of calculation steps. 
We will present a systematical procedure to identify relevant submatrices in the following.

\subsection{Relevant set of submatrices} 

\subsubsection{For $J=0$ and $k= 1,2,3,4,5$}

Let us suppose that $J=0$. For $k= 1,2,3 $, the matrix inequalities both in the diagonal shift and the spreading shift of Fig.~\ref{fig:v} (a) and (b) hold for any $n \ge 0$. This leads to $\| A _{\{0\}}^{(0)}\| \ge \| A _{\{n\}}^{(0)} \|$ for $k=1$,  $\| A _{\{0,1\}}^{(0)}\| \ge \| A _{\{n, n'\}}^{(0)}\|$ for $k=2$, and  $\| A _{\{0,1,2\}}^{(0)}\| \ge \| A _{\{n, n', n''\}}^{(0)}\|$ for $k=3$.

For $k=4$, we can show the inequality  for the spreading shift 
 $  \|  A^{(0)} _{\{n,n+1,n+2,n+3\} } \| \ge  \|  A^{(0)} _{\{n,n+n',n+n'',n+n'''\} } \|$ by using Property (ii) and the inequalities in the diagonal shift of the $ 2 \times 2$ and $3 \times 3$ matrices above. From the row of $l=3$ in Table \ref{table1}, the inequalities for the diagonal shift is fulfilled whenever $n\ge t_l -l =2 $. Hence, the only submatrices that could not be connected by the inequalities are  $A^{(0)}_{\{0,1,2,3\}}$, $A^{(0)}_{\{1,2,3,4\}}$, and  $A^{(0)}_{\{2,3,4,5\}}$. Therefore, the maximum in Eq.~\eqref{Obound} is obtained by comparing the first three matrices, i.e. $\max_{n\in \{ 0, 1,2 \}} \| A^{(0)}_{\{n,n+1,n+2,n+3\}}\| $.

 For $k=5$, we can show the inequality  for the spreading shift 
 $  \|  A^{(0)} _{\{n,n+1,n+2,n+3,n+4\} } \| \ge  \|  A^{(0)} _{\{n,n+n',n+n'',n+n''',n+n''''\} } \|$ by using Property (ii) and the results above except for the case of $n=0$. For $n=0$, we could not have the matrix inequality
  for  $A_{\{0,1,2,3,4\}}^{(0)}$ and $A_{\{0,2,3,4,5\}}^{(0)}$ because
 the $4 \times 4$ matrix inequality could not hold for the first two case of the diagonal shift {($n\ge t_3-3=2$)}.  From Property (i) and the results of $k \le 4$ above, we also have the inequality for the diagonal shift  $  \|  A^{(0)} _{\{n,n+1,n+2,n+3,n+4\} } \| \ge  \|  A^{(0)} _{\{n+1,n+2,n+3,n+4,n+5\} } \|$ when $n \ge t_4 -4 = 5$. In this case,  the matrix inequality for the diagonal shift could not hold for the first five $5 \times 5$ submatrices $\| A_{\{n,n+1,n+2,n+3,n+4\}}\| $ with  ${n \in   \{ 0, 1,2,3,4 \}} $.  Therefore, the optimization can be done by taking the largest one of  $\| A_{\{0,2,3,4,5\}}\|$ and  $\| A_{\{n,n+1,n+2,n+3,n+4\}}\| $ with  ${n \in  \{ 0, 1,2,3,4,5 \}} $.

\subsubsection{For general $J$ and $k$} \label{sis-process}

For temporary simplicity,  let us suppose $|J|\le 4$ [It corresponds to $u ^{(J)} = 0$ in Eq.~(\ref{breve})].
The set of submatrices which could not be connected by the inequalities  for given $k$ can be  specified from $t_{k-1},t_{k-2}, \cdots, t_0$ in Table \ref{table1} as follows: 
First we generate the number of $t_{k-1}-(k-1) +1 =t_{k-1}-k+ 2 $ sets of $\vec n$ 
 in which the submatrix corresponding to $A_{\vec n}$  could not be connected by the inequalities with respect to the diagonal shift:  
{\footnotesize
\begin{eqnarray}
&&\{0, 1, \cdots, k-1\}, \nonumber\\
&&\{ 1,2, \cdots, k-1,k\}, \nonumber\\
&& \ \ \vdots\nonumber\\
&&\{t_{k-1}-k, \cdots, t_{k-1}-1\},  \nonumber\\
&&\{t_{k-1}-k+1, \cdots, t_{k-1}\}.  \label{setelemenets}
\end{eqnarray}}

Second, we generate the sets by repeating the diagonal shift of the last $k-1$ elements of each set of Eq.~(\ref{setelemenets}) until the last index  of $\vec n$ fulfills $n_k > t_{k-2}$ as 
{\footnotesize
\begin{flalign}
 & \{0, \underbrace{1, \cdots, k-1}_{k-1\ elements }\},\{0, \underbrace{2,3, \cdots, k}_{\to \rm Shifted    }\}, \{0, \underbrace{3,4, \cdots, k+1}_{\to  \rm Shifted  }\}, \cdots \nonumber \\
& \{1,\underbrace{2, \cdots, k-1,k}_{k-1\ \rm elements  }\}, \{ 1, \underbrace{3,4, \cdots, k-1,k}_{\to \rm Shifted  }\},\{ 1, \underbrace{4,5, \cdots, k,k+1}_{\to  \rm Shifted   }\}, \cdots \nonumber\\
&  \vdots\nonumber\\
& \{ t_{k-1}\!-\!k, \cdots,  t_{k-1} \! -\! 1\},\{ t_{k-1}\!-\!k, t_{k-1}-k+2, \cdots,  t_{k-1}\},  \cdots \nonumber\\
&\{ t_{k-1}\!-\!k\!+\!1, \cdots,  t_{k-1}\}, \{ t_{k-1}-\!k\!+\!1,t_{k-1}\!-\!k\!+\!3, \cdots,  t_{k\!-\!1}\!+\!1\}, \cdots \nonumber\\  \label{ts}
\end{flalign} }

Third and finally,  we generate the sets by repeating the diagonal shift of the last $k-2$ elements of each set of Eq.~(\ref{ts})  until the last index  of $\vec n$ fulfills $n_k > t_{k-3}$. For example from the elements in the first line of Eq.~(\ref{ts}) we have
{\footnotesize 
\begin{flalign}
&  \{0,1, \underbrace{2, \cdots, k-1}_{k-2\ \rm elements }\}, \{0,1, \underbrace{3,4, \cdots, k}_{ \to \rm Shifted  }\}, \{0,1,  \underbrace{4,5, \cdots, k+1}_{ \to \rm Shifted   } \},\cdots    \nonumber\\
&  \{0,2, \underbrace{3,4, \cdots, k}_{k-2\ \rm elements  }\},  \{0,2, \underbrace{4,5, \cdots, k+1}_{ \to \rm Shifted }\}, \{0,2,  \underbrace{5,6, \cdots, k+2}_{ \to \rm Shifted   } \},\cdots    \nonumber\\
 &  \{0, 3, \underbrace{4,5, \cdots, k+1}_{k-2\ \rm elements  }\},  \{0,3, \underbrace{5,6, \cdots, k+2}_{ \to \rm  Shifted }\}, \cdots  \nonumber\\  & \vdots
\label{tsss}
\end{flalign}}In this manner, we can obtain the total number of, at most, $\prod_{l=0}^{k-1}(t_l - l -1)$ sets of indices $ \vec n $.

For the case of $|J| >4$, we modify the generation process by using $u ^{(J)}$ of Eq.~(\ref{breve}) so that the diagonal shift of the last $l $ elements is repeated 
 until the last index of $\vec n$ fulfills $n_k > \max \{ t_{l-1}, u^{(J)}  + l-1 \}$.

\subsection{Outline for numerical calculation}
Suppose that  $k$, $\kappa$ and $x$ are given.
We first search the set of $\{A^{(J)}\}_J$ which include $k \times k$ principle submatrices whose trace is grater than the conjectured maximum value $\|A_{\{0,1,2, \cdots, k-1 \}}^{(0)}\|$. This process can be executed by  only using the diagonal elements of   $\{A^{(J)}\}_J$.

Next,  we determine the relevant submatrices according to the process described in Appendix~\ref{sis-process} for relevant $J$.

Lastly, the maximum eigenvalues are directly compared to determine the maximum. 


\begin{thebibliography}{51}%
\makeatletter
\providecommand \@ifxundefined [1]{%
 \@ifx{#1\undefined}
}%
\providecommand \@ifnum [1]{%
 \ifnum #1\expandafter \@firstoftwo
 \else \expandafter \@secondoftwo
 \fi
}%
\providecommand \@ifx [1]{%
 \ifx #1\expandafter \@firstoftwo
 \else \expandafter \@secondoftwo
 \fi
}%
\providecommand \natexlab [1]{#1}%
\providecommand \enquote  [1]{``#1''}%
\providecommand \bibnamefont  [1]{#1}%
\providecommand \bibfnamefont [1]{#1}%
\providecommand \citenamefont [1]{#1}%
\providecommand \href@noop [0]{\@secondoftwo}%
\providecommand \href [0]{\begingroup \@sanitize@url \@href}%
\providecommand \@href[1]{\@@startlink{#1}\@@href}%
\providecommand \@@href[1]{\endgroup#1\@@endlink}%
\providecommand \@sanitize@url [0]{\catcode `\\12\catcode `\$12\catcode
  `\&12\catcode `\#12\catcode `\^12\catcode `\_12\catcode `\%12\relax}%
\providecommand \@@startlink[1]{}%
\providecommand \@@endlink[0]{}%
\providecommand \url  [0]{\begingroup\@sanitize@url \@url }%
\providecommand \@url [1]{\endgroup\@href {#1}{\urlprefix }}%
\providecommand \urlprefix  [0]{URL }%
\providecommand \Eprint [0]{\href }%
\providecommand \doibase [0]{http://dx.doi.org/}%
\providecommand \selectlanguage [0]{\@gobble}%
\providecommand \bibinfo  [0]{\@secondoftwo}%
\providecommand \bibfield  [0]{\@secondoftwo}%
\providecommand \translation [1]{[#1]}%
\providecommand \BibitemOpen [0]{}%
\providecommand \bibitemStop [0]{}%
\providecommand \bibitemNoStop [0]{.\EOS\space}%
\providecommand \EOS [0]{\spacefactor3000\relax}%
\providecommand \BibitemShut  [1]{\csname bibitem#1\endcsname}%
\let\auto@bib@innerbib\@empty
\bibitem [{\citenamefont {Horodecki}\ \emph {et~al.}(2009)\citenamefont
  {Horodecki}, \citenamefont {Horodecki}, \citenamefont {Horodecki},\ and\
  \citenamefont {Horodecki}}]{Horo09}%
  \BibitemOpen
  \bibfield  {author} {\bibinfo {author} {\bibfnamefont {R.}~\bibnamefont
  {Horodecki}}, \bibinfo {author} {\bibfnamefont {P.}~\bibnamefont
  {Horodecki}}, \bibinfo {author} {\bibfnamefont {M.}~\bibnamefont
  {Horodecki}}, \ and\ \bibinfo {author} {\bibfnamefont {K.}~\bibnamefont
  {Horodecki}},\ }\bibfield  {title} {\enquote {\bibinfo {title} {Quantum
  entanglement},}\ }\href {\doibase 10.1103/RevModPhys.81.865} {\bibfield
  {journal} {\bibinfo  {journal} {Rev. Mod. Phys.}\ }\textbf {\bibinfo {volume}
  {81}},\ \bibinfo {pages} {865--942} (\bibinfo {year} {2009})}\BibitemShut
  {NoStop}%
\bibitem [{\citenamefont {Pan}\ \emph {et~al.}(2012)\citenamefont {Pan},
  \citenamefont {Chen}, \citenamefont {Lu}, \citenamefont {Weinfurter},
  \citenamefont {Zeilinger},\ and\ \citenamefont {\ifmmode~\dot{Z}\else
  \.{Z}\fi{}ukowski}}]{Pan12}%
  \BibitemOpen
  \bibfield  {author} {\bibinfo {author} {\bibfnamefont {J.-W.}\ \bibnamefont
  {Pan}}, \bibinfo {author} {\bibfnamefont {Z.-B.}\ \bibnamefont {Chen}},
  \bibinfo {author} {\bibfnamefont {C.-Y.}\ \bibnamefont {Lu}}, \bibinfo
  {author} {\bibfnamefont {H.}~\bibnamefont {Weinfurter}}, \bibinfo {author}
  {\bibfnamefont {A.}~\bibnamefont {Zeilinger}}, \ and\ \bibinfo {author}
  {\bibfnamefont {M.}~\bibnamefont {\ifmmode~\dot{Z}\else \.{Z}\fi{}ukowski}},\
  }\bibfield  {title} {\enquote {\bibinfo {title} {Multiphoton entanglement and
  interferometry},}\ }\href {\doibase 10.1103/RevModPhys.84.777} {\bibfield
  {journal} {\bibinfo  {journal} {Rev. Mod. Phys.}\ }\textbf {\bibinfo {volume}
  {84}},\ \bibinfo {pages} {777--838} (\bibinfo {year} {2012})}\BibitemShut
  {NoStop}%
\bibitem [{\citenamefont {Terhal}\ and\ \citenamefont
  {Horodecki}(2000)}]{Ter20}%
  \BibitemOpen
  \bibfield  {author} {\bibinfo {author} {\bibfnamefont {B.M.}\ \bibnamefont
  {Terhal}}\ and\ \bibinfo {author} {\bibfnamefont {P.}~\bibnamefont
  {Horodecki}},\ }\bibfield  {title} {\enquote {\bibinfo {title} {Schmidt
  number for density matrices},}\ }\href {\doibase 10.1103/PhysRevA.61.040301}
  {\bibfield  {journal} {\bibinfo  {journal} {Phys. Rev. A}\ }\textbf {\bibinfo
  {volume} {61}},\ \bibinfo {pages} {040301} (\bibinfo {year}
  {2000})}\BibitemShut {NoStop}%
\bibitem [{\citenamefont {Huang}(2006)}]{Hua06}%
  \BibitemOpen
  \bibfield  {author} {\bibinfo {author} {\bibfnamefont {S.}~\bibnamefont
  {Huang}},\ }\bibfield  {title} {\enquote {\bibinfo {title} {Schmidt number
  for quantum operations},}\ }\href {\doibase 10.1103/PhysRevA.73.052318}
  {\bibfield  {journal} {\bibinfo  {journal} {Phys. Rev. A}\ }\textbf {\bibinfo
  {volume} {73}},\ \bibinfo {pages} {052318} (\bibinfo {year}
  {2006})}\BibitemShut {NoStop}%
\bibitem [{\citenamefont {Chruściński}\ and\ \citenamefont
  {Kossakowski}(2006)}]{Chru06}%
  \BibitemOpen
  \bibfield  {author} {\bibinfo {author} {\bibfnamefont {Dariusz}\ \bibnamefont
  {Chruściński}}\ and\ \bibinfo {author} {\bibfnamefont {Andrzej}\
  \bibnamefont {Kossakowski}},\ }\bibfield  {title} {\enquote {\bibinfo {title}
  {On partially entanglement breaking channels},}\ }\href {\doibase
  10.1007/s11080-006-7264-7} {\bibfield  {journal} {\bibinfo  {journal} {Open
  Systems \& Information Dynamics}\ }\textbf {\bibinfo {volume} {13}},\
  \bibinfo {pages} {17--26} (\bibinfo {year} {2006})}\BibitemShut {NoStop}%
\bibitem [{\citenamefont {Namiki}(2013)}]{Namiki13a}%
  \BibitemOpen
  \bibfield  {author} {\bibinfo {author} {\bibfnamefont {R.}~\bibnamefont
  {Namiki}},\ }\bibfield  {title} {\enquote {\bibinfo {title} {Composability of
  partial-entanglement-breaking channels via entanglement-assisted local
  operations and classical communication},}\ }\href@noop {} {\bibfield
  {journal} {\bibinfo  {journal} {Phys. Rev. A}\ }\textbf {\bibinfo {volume}
  {88}},\ \bibinfo {pages} {064301} (\bibinfo {year} {2013})}\BibitemShut
  {NoStop}%
\bibitem [{\citenamefont {Horodecki}\ \emph {et~al.}(2003)\citenamefont
  {Horodecki}, \citenamefont {Shor},\ and\ \citenamefont {Ruskai}}]{Horo03a}%
  \BibitemOpen
  \bibfield  {author} {\bibinfo {author} {\bibfnamefont {M.}~\bibnamefont
  {Horodecki}}, \bibinfo {author} {\bibfnamefont {P.~W.}\ \bibnamefont {Shor}},
  \ and\ \bibinfo {author} {\bibfnamefont {M.~B.}\ \bibnamefont {Ruskai}},\
  }\bibfield  {title} {\enquote {\bibinfo {title} {{Entanglement Breaking
  Channels}},}\ }\href@noop {} {\bibfield  {journal} {\bibinfo  {journal} {Rev.
  Math. Phys.}\ }\textbf {\bibinfo {volume} {15}},\ \bibinfo {pages} {629}
  (\bibinfo {year} {2003})}\BibitemShut {NoStop}%
\bibitem [{\citenamefont {Holevo}(2008)}]{Hol08}%
  \BibitemOpen
  \bibfield  {author} {\bibinfo {author} {\bibfnamefont {A.~S.}\ \bibnamefont
  {Holevo}},\ }\bibfield  {title} {\enquote {\bibinfo {title}
  {{Entanglement-breaking channels in infinite dimensions}},}\ }\href {\doibase
  10.1134/S0032946008030010} {\bibfield  {journal} {\bibinfo  {journal} {Probl.
  Info. Transm.}\ }\textbf {\bibinfo {volume} {44}},\ \bibinfo {pages}
  {171--184} (\bibinfo {year} {2008})}\BibitemShut {NoStop}%
\bibitem [{\citenamefont {Namiki}\ and\ \citenamefont
  {Tokunaga}(2012{\natexlab{a}})}]{Namiki12R}%
  \BibitemOpen
  \bibfield  {author} {\bibinfo {author} {\bibfnamefont {R.}~\bibnamefont
  {Namiki}}\ and\ \bibinfo {author} {\bibfnamefont {Y.}~\bibnamefont
  {Tokunaga}},\ }\bibfield  {title} {\enquote {\bibinfo {title}
  {{Schmidt-number benchmark for genuine quantum memories and gates}},}\ }\href
  {\doibase 10.1103/PhysRevA.85.010305} {\bibfield  {journal} {\bibinfo
  {journal} {Phys. Rev. A}\ }\textbf {\bibinfo {volume} {85}},\ \bibinfo
  {pages} {010305(R)} (\bibinfo {year} {2012}{\natexlab{a}})}\BibitemShut
  {NoStop}%
\bibitem [{\citenamefont {Sanpera}\ \emph {et~al.}(2001)\citenamefont
  {Sanpera}, \citenamefont {Bru\ss{}},\ and\ \citenamefont
  {Lewenstein}}]{Sanpera01}%
  \BibitemOpen
  \bibfield  {author} {\bibinfo {author} {\bibfnamefont {A.}~\bibnamefont
  {Sanpera}}, \bibinfo {author} {\bibfnamefont {D.}~\bibnamefont {Bru\ss{}}}, \
  and\ \bibinfo {author} {\bibfnamefont {M.}~\bibnamefont {Lewenstein}},\
  }\bibfield  {title} {\enquote {\bibinfo {title} {Schmidt-number witnesses and
  bound entanglement},}\ }\href {\doibase 10.1103/PhysRevA.63.050301}
  {\bibfield  {journal} {\bibinfo  {journal} {Phys. Rev. A}\ }\textbf {\bibinfo
  {volume} {63}},\ \bibinfo {pages} {050301} (\bibinfo {year}
  {2001})}\BibitemShut {NoStop}%
\bibitem [{\citenamefont {Tokunaga}\ \emph {et~al.}(2006)\citenamefont
  {Tokunaga}, \citenamefont {Yamamoto}, \citenamefont {Koashi},\ and\
  \citenamefont {Imoto}}]{Tokunaga06}%
  \BibitemOpen
  \bibfield  {author} {\bibinfo {author} {\bibfnamefont {Y.}~\bibnamefont
  {Tokunaga}}, \bibinfo {author} {\bibfnamefont {T.}~\bibnamefont {Yamamoto}},
  \bibinfo {author} {\bibfnamefont {M.}~\bibnamefont {Koashi}}, \ and\ \bibinfo
  {author} {\bibfnamefont {N.}~\bibnamefont {Imoto}},\ }\bibfield  {title}
  {\enquote {\bibinfo {title} {Fidelity estimation and entanglement
  verification for experimentally produced four-qubit cluster states},}\ }\href
  {\doibase 10.1103/PhysRevA.74.020301} {\bibfield  {journal} {\bibinfo
  {journal} {Phys. Rev. A}\ }\textbf {\bibinfo {volume} {74}},\ \bibinfo
  {pages} {020301(R)} (\bibinfo {year} {2006})}\BibitemShut {NoStop}%
\bibitem [{\citenamefont {Tokunaga}\ \emph {et~al.}(2008)\citenamefont
  {Tokunaga}, \citenamefont {Kuwashiro}, \citenamefont {Yamamoto},
  \citenamefont {Koashi},\ and\ \citenamefont {Imoto}}]{Tokunaga08}%
  \BibitemOpen
  \bibfield  {author} {\bibinfo {author} {\bibfnamefont {Y.}~\bibnamefont
  {Tokunaga}}, \bibinfo {author} {\bibfnamefont {S.}~\bibnamefont {Kuwashiro}},
  \bibinfo {author} {\bibfnamefont {T.}~\bibnamefont {Yamamoto}}, \bibinfo
  {author} {\bibfnamefont {M.}~\bibnamefont {Koashi}}, \ and\ \bibinfo {author}
  {\bibfnamefont {N.}~\bibnamefont {Imoto}},\ }\bibfield  {title} {\enquote
  {\bibinfo {title} {Generation of high-fidelity four-photon cluster state and
  quantum-domain demonstration of one-way quantum computing},}\ }\href
  {\doibase 10.1103/PhysRevLett.100.210501} {\bibfield  {journal} {\bibinfo
  {journal} {Phys. Rev. Lett.}\ }\textbf {\bibinfo {volume} {100}},\ \bibinfo
  {pages} {210501} (\bibinfo {year} {2008})}\BibitemShut {NoStop}%
\bibitem [{\citenamefont {Inoue}\ \emph {et~al.}(2009)\citenamefont {Inoue},
  \citenamefont {Yonehara}, \citenamefont {Miyamoto}, \citenamefont {Koashi},\
  and\ \citenamefont {Kozuma}}]{Inoue09}%
  \BibitemOpen
  \bibfield  {author} {\bibinfo {author} {\bibfnamefont {R.}~\bibnamefont
  {Inoue}}, \bibinfo {author} {\bibfnamefont {T.}~\bibnamefont {Yonehara}},
  \bibinfo {author} {\bibfnamefont {Y.}~\bibnamefont {Miyamoto}}, \bibinfo
  {author} {\bibfnamefont {M.}~\bibnamefont {Koashi}}, \ and\ \bibinfo {author}
  {\bibfnamefont {M.}~\bibnamefont {Kozuma}},\ }\bibfield  {title} {\enquote
  {\bibinfo {title} {Measuring qutrit-qutrit entanglement of orbital angular
  momentum states of an atomic ensemble and a photon},}\ }\href {\doibase
  10.1103/PhysRevLett.103.110503} {\bibfield  {journal} {\bibinfo  {journal}
  {Phys. Rev. Lett.}\ }\textbf {\bibinfo {volume} {103}},\ \bibinfo {pages}
  {110503} (\bibinfo {year} {2009})}\BibitemShut {NoStop}%
\bibitem [{\citenamefont {Li}\ \emph {et~al.}(2010)\citenamefont {Li},
  \citenamefont {Chen}, \citenamefont {Reingruber}, \citenamefont {Chen},\ and\
  \citenamefont {Pan}}]{Li10}%
  \BibitemOpen
  \bibfield  {author} {\bibinfo {author} {\bibfnamefont {C.-M.}\ \bibnamefont
  {Li}}, \bibinfo {author} {\bibfnamefont {K.}~\bibnamefont {Chen}}, \bibinfo
  {author} {\bibfnamefont {A.}~\bibnamefont {Reingruber}}, \bibinfo {author}
  {\bibfnamefont {Y.-N.}\ \bibnamefont {Chen}}, \ and\ \bibinfo {author}
  {\bibfnamefont {J.-W.}\ \bibnamefont {Pan}},\ }\bibfield  {title} {\enquote
  {\bibinfo {title} {Verifying genuine high-order entanglement},}\ }\href
  {\doibase 10.1103/PhysRevLett.105.210504} {\bibfield  {journal} {\bibinfo
  {journal} {Phys. Rev. Lett.}\ }\textbf {\bibinfo {volume} {105}},\ \bibinfo
  {pages} {210504} (\bibinfo {year} {2010})}\BibitemShut {NoStop}%
\bibitem [{\citenamefont {Sperling}\ and\ \citenamefont
  {Vogel}(2011)}]{Sperl11}%
  \BibitemOpen
  \bibfield  {author} {\bibinfo {author} {\bibfnamefont {J.}~\bibnamefont
  {Sperling}}\ and\ \bibinfo {author} {\bibfnamefont {W.}~\bibnamefont
  {Vogel}},\ }\bibfield  {title} {\enquote {\bibinfo {title} {Determination of
  the schmidt number},}\ }\href {\doibase 10.1103/PhysRevA.83.042315}
  {\bibfield  {journal} {\bibinfo  {journal} {Phys. Rev. A}\ }\textbf {\bibinfo
  {volume} {83}},\ \bibinfo {pages} {042315} (\bibinfo {year}
  {2011})}\BibitemShut {NoStop}%
\bibitem [{\citenamefont {Namiki}\ and\ \citenamefont
  {Tokunaga}(2012{\natexlab{b}})}]{Namiki12L}%
  \BibitemOpen
  \bibfield  {author} {\bibinfo {author} {\bibfnamefont {R.}~\bibnamefont
  {Namiki}}\ and\ \bibinfo {author} {\bibfnamefont {Y.}~\bibnamefont
  {Tokunaga}},\ }\bibfield  {title} {\enquote {\bibinfo {title} {Discrete
  fourier-based correlations for entanglement detection},}\ }\href {\doibase
  10.1103/PhysRevLett.108.230503} {\bibfield  {journal} {\bibinfo  {journal}
  {Phys. Rev. Lett.}\ }\textbf {\bibinfo {volume} {108}},\ \bibinfo {pages}
  {230503} (\bibinfo {year} {2012}{\natexlab{b}})}\BibitemShut {NoStop}%
\bibitem [{\citenamefont {Shahandeh}\ \emph {et~al.}(2013)\citenamefont
  {Shahandeh}, \citenamefont {Sperling},\ and\ \citenamefont
  {Vogel}}]{Shahandeh2013}%
  \BibitemOpen
  \bibfield  {author} {\bibinfo {author} {\bibfnamefont {F.}~\bibnamefont
  {Shahandeh}}, \bibinfo {author} {\bibfnamefont {J.}~\bibnamefont {Sperling}},
  \ and\ \bibinfo {author} {\bibfnamefont {W.}~\bibnamefont {Vogel}},\
  }\bibfield  {title} {\enquote {\bibinfo {title} {{Operational Gaussian
  Schmidt-number witnesses}},}\ }\href {\doibase 10.1103/PhysRevA.88.062323}
  {\bibfield  {journal} {\bibinfo  {journal} {Phys. Rev. A}\ }\textbf {\bibinfo
  {volume} {88}},\ \bibinfo {pages} {062323} (\bibinfo {year}
  {2013})}\BibitemShut {NoStop}%
\bibitem [{\citenamefont {Guti\'errez-Esparza}\ \emph
  {et~al.}(2014)\citenamefont {Guti\'errez-Esparza}, \citenamefont {Pimenta},
  \citenamefont {Marques}, \citenamefont {Matoso}, \citenamefont {Sperling},
  \citenamefont {Vogel},\ and\ \citenamefont {P\'adua}}]{Guti14}%
  \BibitemOpen
  \bibfield  {author} {\bibinfo {author} {\bibfnamefont {A.~J.}\ \bibnamefont
  {Guti\'errez-Esparza}}, \bibinfo {author} {\bibfnamefont {W.~M.}\
  \bibnamefont {Pimenta}}, \bibinfo {author} {\bibfnamefont {B.}~\bibnamefont
  {Marques}}, \bibinfo {author} {\bibfnamefont {A.~A.}\ \bibnamefont {Matoso}},
  \bibinfo {author} {\bibfnamefont {J.}~\bibnamefont {Sperling}}, \bibinfo
  {author} {\bibfnamefont {W.}~\bibnamefont {Vogel}}, \ and\ \bibinfo {author}
  {\bibfnamefont {S.}~\bibnamefont {P\'adua}},\ }\bibfield  {title} {\enquote
  {\bibinfo {title} {Detection of nonlocal superpositions},}\ }\href {\doibase
  10.1103/PhysRevA.90.032328} {\bibfield  {journal} {\bibinfo  {journal} {Phys.
  Rev. A}\ }\textbf {\bibinfo {volume} {90}},\ \bibinfo {pages} {032328}
  (\bibinfo {year} {2014})}\BibitemShut {NoStop}%
\bibitem [{\citenamefont {Braunstein}\ and\ \citenamefont {van
  Loock}(2005)}]{Bra05}%
  \BibitemOpen
  \bibfield  {author} {\bibinfo {author} {\bibfnamefont {S.L.}\ \bibnamefont
  {Braunstein}}\ and\ \bibinfo {author} {\bibfnamefont {P.}~\bibnamefont {van
  Loock}},\ }\bibfield  {title} {\enquote {\bibinfo {title} {Quantum
  information with continuous variables},}\ }\href {\doibase
  10.1103/RevModPhys.77.513} {\bibfield  {journal} {\bibinfo  {journal} {Rev.
  Mod. Phys.}\ }\textbf {\bibinfo {volume} {77}},\ \bibinfo {pages} {513--577}
  (\bibinfo {year} {2005})}\BibitemShut {NoStop}%
\bibitem [{\citenamefont {Hammerer}\ \emph {et~al.}(2010)\citenamefont
  {Hammerer}, \citenamefont {S\o{}rensen},\ and\ \citenamefont
  {Polzik}}]{Hamm10}%
  \BibitemOpen
  \bibfield  {author} {\bibinfo {author} {\bibfnamefont {K.}~\bibnamefont
  {Hammerer}}, \bibinfo {author} {\bibfnamefont {A.S.}\ \bibnamefont
  {S\o{}rensen}}, \ and\ \bibinfo {author} {\bibfnamefont {E.S.}\ \bibnamefont
  {Polzik}},\ }\bibfield  {title} {\enquote {\bibinfo {title} {Quantum
  interface between light and atomic ensembles},}\ }\href {\doibase
  10.1103/RevModPhys.82.1041} {\bibfield  {journal} {\bibinfo  {journal} {Rev.
  Mod. Phys.}\ }\textbf {\bibinfo {volume} {82}},\ \bibinfo {pages}
  {1041--1093} (\bibinfo {year} {2010})}\BibitemShut {NoStop}%
\bibitem [{\citenamefont {Weedbrook}\ \emph {et~al.}(2012)\citenamefont
  {Weedbrook}, \citenamefont {Pirandola}, \citenamefont {Garc\'{\i}a-Patr\'on},
  \citenamefont {Cerf}, \citenamefont {Ralph}, \citenamefont {Shapiro},\ and\
  \citenamefont {Lloyd}}]{Weed12}%
  \BibitemOpen
  \bibfield  {author} {\bibinfo {author} {\bibfnamefont {C.}~\bibnamefont
  {Weedbrook}}, \bibinfo {author} {\bibfnamefont {S.}~\bibnamefont
  {Pirandola}}, \bibinfo {author} {\bibfnamefont {R.}~\bibnamefont
  {Garc\'{\i}a-Patr\'on}}, \bibinfo {author} {\bibfnamefont {N.J.}\
  \bibnamefont {Cerf}}, \bibinfo {author} {\bibfnamefont {T.C.}\ \bibnamefont
  {Ralph}}, \bibinfo {author} {\bibfnamefont {J.H.}\ \bibnamefont {Shapiro}}, \
  and\ \bibinfo {author} {\bibfnamefont {S.}~\bibnamefont {Lloyd}},\ }\bibfield
   {title} {\enquote {\bibinfo {title} {Gaussian quantum information},}\ }\href
  {\doibase 10.1103/RevModPhys.84.621} {\bibfield  {journal} {\bibinfo
  {journal} {Rev. Mod. Phys.}\ }\textbf {\bibinfo {volume} {84}},\ \bibinfo
  {pages} {621--669} (\bibinfo {year} {2012})}\BibitemShut {NoStop}%
\bibitem [{\citenamefont {Furusawa}\ \emph {et~al.}(1998)\citenamefont
  {Furusawa}, \citenamefont {S{\o}rensen}, \citenamefont {Braunstein},
  \citenamefont {Fuchs}, \citenamefont {Kimble},\ and\ \citenamefont
  {Polzik}}]{Furusawa98}%
  \BibitemOpen
  \bibfield  {author} {\bibinfo {author} {\bibfnamefont {A.}~\bibnamefont
  {Furusawa}}, \bibinfo {author} {\bibfnamefont {J.~L.}\ \bibnamefont
  {S{\o}rensen}}, \bibinfo {author} {\bibfnamefont {S.~L.}\ \bibnamefont
  {Braunstein}}, \bibinfo {author} {\bibfnamefont {C.~A.}\ \bibnamefont
  {Fuchs}}, \bibinfo {author} {\bibfnamefont {H.~J.}\ \bibnamefont {Kimble}}, \
  and\ \bibinfo {author} {\bibfnamefont {E.~S.}\ \bibnamefont {Polzik}},\
  }\bibfield  {title} {\enquote {\bibinfo {title} {{Unconditional quantum
  teleportation}},}\ }\href
  {http://www.sciencemag.org/content/282/5389/706.short} {\bibfield  {journal}
  {\bibinfo  {journal} {Science}\ }\textbf {\bibinfo {volume} {282}},\ \bibinfo
  {pages} {706--710} (\bibinfo {year} {1998})}\BibitemShut {NoStop}%
\bibitem [{\citenamefont {Julsgaard}\ \emph {et~al.}(2004)\citenamefont
  {Julsgaard}, \citenamefont {Sherson}, \citenamefont {Cirac}, \citenamefont
  {Fiurasek},\ and\ \citenamefont {Polzik}}]{julsgaard04a}%
  \BibitemOpen
  \bibfield  {author} {\bibinfo {author} {\bibfnamefont {Brian}\ \bibnamefont
  {Julsgaard}}, \bibinfo {author} {\bibfnamefont {Jacob}\ \bibnamefont
  {Sherson}}, \bibinfo {author} {\bibfnamefont {J.\~{}Ignacio}\ \bibnamefont
  {Cirac}}, \bibinfo {author} {\bibfnamefont {Jaromir}\ \bibnamefont
  {Fiurasek}}, \ and\ \bibinfo {author} {\bibfnamefont {Eugene~S}\ \bibnamefont
  {Polzik}},\ }\bibfield  {title} {\enquote {\bibinfo {title} {{Experimental
  demonstration of quantum memory for light}},}\ }\href@noop {} {\bibfield
  {journal} {\bibinfo  {journal} {Nature}\ }\textbf {\bibinfo {volume} {432}},\
  \bibinfo {pages} {482} (\bibinfo {year} {2004})}\BibitemShut {NoStop}%
\bibitem [{\citenamefont {Lobino}\ \emph {et~al.}(2009)\citenamefont {Lobino},
  \citenamefont {Kupchak}, \citenamefont {Figueroa},\ and\ \citenamefont
  {Lvovsky}}]{Lob09}%
  \BibitemOpen
  \bibfield  {author} {\bibinfo {author} {\bibfnamefont {M.}~\bibnamefont
  {Lobino}}, \bibinfo {author} {\bibfnamefont {C.}~\bibnamefont {Kupchak}},
  \bibinfo {author} {\bibfnamefont {E.}~\bibnamefont {Figueroa}}, \ and\
  \bibinfo {author} {\bibfnamefont {A.~I.}\ \bibnamefont {Lvovsky}},\
  }\bibfield  {title} {\enquote {\bibinfo {title} {Memory for light as a
  quantum process},}\ }\href {\doibase 10.1103/PhysRevLett.102.203601}
  {\bibfield  {journal} {\bibinfo  {journal} {Phys. Rev. Lett.}\ }\textbf
  {\bibinfo {volume} {102}},\ \bibinfo {pages} {203601} (\bibinfo {year}
  {2009})}\BibitemShut {NoStop}%
\bibitem [{\citenamefont {Braunstein}\ \emph {et~al.}(2000)\citenamefont
  {Braunstein}, \citenamefont {Fuchs},\ and\ \citenamefont {Kimble}}]{Bra00}%
  \BibitemOpen
  \bibfield  {author} {\bibinfo {author} {\bibfnamefont {S.L.}\ \bibnamefont
  {Braunstein}}, \bibinfo {author} {\bibfnamefont {C.A.}\ \bibnamefont
  {Fuchs}}, \ and\ \bibinfo {author} {\bibfnamefont {H.J.}\ \bibnamefont
  {Kimble}},\ }\bibfield  {title} {\enquote {\bibinfo {title} {{Criteria for
  continuous-variable quantum teleportation}},}\ }\href
  {http://www.tandfonline.com/doi/abs/10.1080/09500340008244041} {\bibfield
  {journal} {\bibinfo  {journal} {J. Mod. Opt.}\ }\textbf {\bibinfo {volume}
  {47}},\ \bibinfo {pages} {267} (\bibinfo {year} {2000})}\BibitemShut
  {NoStop}%
\bibitem [{\citenamefont {Hammerer}\ \emph {et~al.}(2005)\citenamefont
  {Hammerer}, \citenamefont {Wolf}, \citenamefont {Polzik},\ and\ \citenamefont
  {Cirac}}]{Ham05}%
  \BibitemOpen
  \bibfield  {author} {\bibinfo {author} {\bibfnamefont {K.}~\bibnamefont
  {Hammerer}}, \bibinfo {author} {\bibfnamefont {M.~M.}\ \bibnamefont {Wolf}},
  \bibinfo {author} {\bibfnamefont {E.~S.}\ \bibnamefont {Polzik}}, \ and\
  \bibinfo {author} {\bibfnamefont {J.~I.}\ \bibnamefont {Cirac}},\ }\bibfield
  {title} {\enquote {\bibinfo {title} {Quantum benchmark for storage and
  transmission of coherent states},}\ }\href {\doibase
  10.1103/PhysRevLett.94.150503} {\bibfield  {journal} {\bibinfo  {journal}
  {Phys. Rev. Lett.}\ }\textbf {\bibinfo {volume} {94}},\ \bibinfo {pages}
  {150503} (\bibinfo {year} {2005})}\BibitemShut {NoStop}%
\bibitem [{\citenamefont {Namiki}\ \emph {et~al.}(2008)\citenamefont {Namiki},
  \citenamefont {Koashi},\ and\ \citenamefont {Imoto}}]{Namiki07}%
  \BibitemOpen
  \bibfield  {author} {\bibinfo {author} {\bibfnamefont {R.}~\bibnamefont
  {Namiki}}, \bibinfo {author} {\bibfnamefont {M.}~\bibnamefont {Koashi}}, \
  and\ \bibinfo {author} {\bibfnamefont {N.}~\bibnamefont {Imoto}},\ }\bibfield
   {title} {\enquote {\bibinfo {title} {{Fidelity criterion for quantum-domain
  transmission and storage of coherent states beyond the unit-gain
  constraint.}}}\ }\href@noop {} {\bibfield  {journal} {\bibinfo  {journal}
  {Phys. Rev. Lett.}\ }\textbf {\bibinfo {volume} {101}},\ \bibinfo {pages}
  {100502} (\bibinfo {year} {2008})}\BibitemShut {NoStop}%
\bibitem [{\citenamefont {Takano}\ \emph {et~al.}(2008)\citenamefont {Takano},
  \citenamefont {Fuyama}, \citenamefont {Namiki},\ and\ \citenamefont
  {Takahashi}}]{Takano08}%
  \BibitemOpen
  \bibfield  {author} {\bibinfo {author} {\bibfnamefont {T.}~\bibnamefont
  {Takano}}, \bibinfo {author} {\bibfnamefont {M.}~\bibnamefont {Fuyama}},
  \bibinfo {author} {\bibfnamefont {R.}~\bibnamefont {Namiki}}, \ and\ \bibinfo
  {author} {\bibfnamefont {Y.}~\bibnamefont {Takahashi}},\ }\bibfield  {title}
  {\enquote {\bibinfo {title} {{Continuous-variable quantum swapping gate
  between light and atoms}},}\ }\href {\doibase 10.1103/PhysRevA.78.010307}
  {\bibfield  {journal} {\bibinfo  {journal} {Phys. Rev. A}\ }\textbf {\bibinfo
  {volume} {78}},\ \bibinfo {pages} {010307(R)} (\bibinfo {year}
  {2008})}\BibitemShut {NoStop}%
\bibitem [{\citenamefont {Namiki}(2011{\natexlab{a}})}]{Namiki11R}%
  \BibitemOpen
  \bibfield  {author} {\bibinfo {author} {\bibfnamefont {R.}~\bibnamefont
  {Namiki}},\ }\bibfield  {title} {\enquote {\bibinfo {title} {Fundamental
  quantum limits on phase-insensitive linear amplification and phase
  conjugation in a practical framework},}\ }\href {\doibase
  10.1103/PhysRevA.83.040302} {\bibfield  {journal} {\bibinfo  {journal} {Phys.
  Rev. A}\ }\textbf {\bibinfo {volume} {83}},\ \bibinfo {pages} {040302}
  (\bibinfo {year} {2011}{\natexlab{a}})}\BibitemShut {NoStop}%
\bibitem [{\citenamefont {Chiribella}\ and\ \citenamefont
  {Xie}(2013)}]{Chir13}%
  \BibitemOpen
  \bibfield  {author} {\bibinfo {author} {\bibfnamefont {G.}~\bibnamefont
  {Chiribella}}\ and\ \bibinfo {author} {\bibfnamefont {J.}~\bibnamefont
  {Xie}},\ }\bibfield  {title} {\enquote {\bibinfo {title} {Optimal design and
  quantum benchmarks for coherent state amplifiers},}\ }\href {\doibase
  10.1103/PhysRevLett.110.213602} {\bibfield  {journal} {\bibinfo  {journal}
  {Phys. Rev. Lett.}\ }\textbf {\bibinfo {volume} {110}},\ \bibinfo {pages}
  {213602} (\bibinfo {year} {2013})}\BibitemShut {NoStop}%
\bibitem [{\citenamefont {Namiki}(2011{\natexlab{b}})}]{Namiki11}%
  \BibitemOpen
  \bibfield  {author} {\bibinfo {author} {\bibfnamefont {R.}~\bibnamefont
  {Namiki}},\ }\bibfield  {title} {\enquote {\bibinfo {title} {{Simple proof of
  the quantum benchmark fidelity for continuous-variable quantum devices}},}\
  }\href {\doibase 10.1103/PhysRevA.83.042323} {\bibfield  {journal} {\bibinfo
  {journal} {Phys. Rev. A}\ }\textbf {\bibinfo {volume} {83}},\ \bibinfo
  {pages} {042323} (\bibinfo {year} {2011}{\natexlab{b}})}\BibitemShut
  {NoStop}%
\bibitem [{\citenamefont {Yang}\ \emph {et~al.}(2014)\citenamefont {Yang},
  \citenamefont {Chiribella},\ and\ \citenamefont {Adesso}}]{Yang14}%
  \BibitemOpen
  \bibfield  {author} {\bibinfo {author} {\bibfnamefont {Y.}~\bibnamefont
  {Yang}}, \bibinfo {author} {\bibfnamefont {G.}~\bibnamefont {Chiribella}}, \
  and\ \bibinfo {author} {\bibfnamefont {G.}~\bibnamefont {Adesso}},\
  }\bibfield  {title} {\enquote {\bibinfo {title} {Certifying quantumness:
  Benchmarks for the optimal processing of generalized coherent and squeezed
  states},}\ }\href {\doibase 10.1103/PhysRevA.90.042319} {\bibfield  {journal}
  {\bibinfo  {journal} {Phys. Rev. A}\ }\textbf {\bibinfo {volume} {90}},\
  \bibinfo {pages} {042319} (\bibinfo {year} {2014})}\BibitemShut {NoStop}%
\bibitem [{\citenamefont {Namiki}()}]{Namiki1503}%
  \BibitemOpen
  \bibfield  {author} {\bibinfo {author} {\bibfnamefont {R.}~\bibnamefont
  {Namiki}},\ }\bibfield  {title} {\enquote {\bibinfo {title} {Converting
  separable conditions to entanglement breaking conditions},}\ }\href@noop {}
  {\ }\Eprint {http://arxiv.org/abs/arXiv:1503.07109} {arXiv:1503.07109}
  \BibitemShut {NoStop}%
\bibitem [{\citenamefont {Fuchs}\ and\ \citenamefont {Sasaki}(2003)}]{Fuc03}%
  \BibitemOpen
  \bibfield  {author} {\bibinfo {author} {\bibfnamefont {C.~A.}\ \bibnamefont
  {Fuchs}}\ and\ \bibinfo {author} {\bibfnamefont {M.}~\bibnamefont {Sasaki}},\
  }\bibfield  {title} {\enquote {\bibinfo {title} {Squeezing quantum
  information through a classical channel: measuring the quantumness of a set
  of quantum states},}\ }\href@noop {} {\bibfield  {journal} {\bibinfo
  {journal} {Quantum Inf. Comput.}\ }\textbf {\bibinfo {volume} {3}},\ \bibinfo
  {pages} {377} (\bibinfo {year} {2003})}\BibitemShut {NoStop}%
\bibitem [{\citenamefont {Namiki}(2008)}]{Namiki08}%
  \BibitemOpen
  \bibfield  {author} {\bibinfo {author} {\bibfnamefont {R.}~\bibnamefont
  {Namiki}},\ }\bibfield  {title} {\enquote {\bibinfo {title} {Verification of
  the quantum-domain process using two nonorthogonal states},}\ }\href
  {\doibase 10.1103/PhysRevA.78.032333} {\bibfield  {journal} {\bibinfo
  {journal} {Phys. Rev. A}\ }\textbf {\bibinfo {volume} {78}},\ \bibinfo
  {pages} {032333} (\bibinfo {year} {2008})}\BibitemShut {NoStop}%
\bibitem [{\citenamefont {H\"aseler}\ \emph {et~al.}(2008)\citenamefont
  {H\"aseler}, \citenamefont {Moroder},\ and\ \citenamefont
  {L\"utkenhaus}}]{Has08}%
  \BibitemOpen
  \bibfield  {author} {\bibinfo {author} {\bibfnamefont {H.}~\bibnamefont
  {H\"aseler}}, \bibinfo {author} {\bibfnamefont {T.}~\bibnamefont {Moroder}},
  \ and\ \bibinfo {author} {\bibfnamefont {N.}~\bibnamefont {L\"utkenhaus}},\
  }\bibfield  {title} {\enquote {\bibinfo {title} {Testing quantum devices:
  Practical entanglement verification in bipartite optical systems},}\ }\href
  {\doibase 10.1103/PhysRevA.77.032303} {\bibfield  {journal} {\bibinfo
  {journal} {Phys. Rev. A}\ }\textbf {\bibinfo {volume} {77}},\ \bibinfo
  {pages} {032303} (\bibinfo {year} {2008})}\BibitemShut {NoStop}%
\bibitem [{\citenamefont {H\"aseler}\ and\ \citenamefont
  {L\"utkenhaus}(2009)}]{Has09}%
  \BibitemOpen
  \bibfield  {author} {\bibinfo {author} {\bibfnamefont {H.}~\bibnamefont
  {H\"aseler}}\ and\ \bibinfo {author} {\bibfnamefont {N.}~\bibnamefont
  {L\"utkenhaus}},\ }\bibfield  {title} {\enquote {\bibinfo {title} {Probing
  the quantumness of channels with mixed states},}\ }\href {\doibase
  10.1103/PhysRevA.80.042304} {\bibfield  {journal} {\bibinfo  {journal} {Phys.
  Rev. A}\ }\textbf {\bibinfo {volume} {80}},\ \bibinfo {pages} {042304}
  (\bibinfo {year} {2009})}\BibitemShut {NoStop}%
\bibitem [{\citenamefont {Owari}\ \emph {et~al.}(2008)\citenamefont {Owari},
  \citenamefont {Plenio}, \citenamefont {Polzik}, \citenamefont {Serafini},\
  and\ \citenamefont {Wolf}}]{Owari08}%
  \BibitemOpen
  \bibfield  {author} {\bibinfo {author} {\bibfnamefont {M.}~\bibnamefont
  {Owari}}, \bibinfo {author} {\bibfnamefont {M.~B.}\ \bibnamefont {Plenio}},
  \bibinfo {author} {\bibfnamefont {E.~S.}\ \bibnamefont {Polzik}}, \bibinfo
  {author} {\bibfnamefont {A.}~\bibnamefont {Serafini}}, \ and\ \bibinfo
  {author} {\bibfnamefont {M.~M.}\ \bibnamefont {Wolf}},\ }\bibfield  {title}
  {\enquote {\bibinfo {title} {{Squeezing the limit: quantum benchmarks for the
  teleportation and storage of squeezed states}},}\ }\href {\doibase
  10.1088/1367-2630/10/11/113014} {\bibfield  {journal} {\bibinfo  {journal}
  {New J. Phys.}\ }\textbf {\bibinfo {volume} {10}},\ \bibinfo {pages} {113014}
  (\bibinfo {year} {2008})}\BibitemShut {NoStop}%
\bibitem [{\citenamefont {Namiki}\ and\ \citenamefont
  {Azuma}(2015)}]{Namiki-Azuma13x}%
  \BibitemOpen
  \bibfield  {author} {\bibinfo {author} {\bibfnamefont {R.}~\bibnamefont
  {Namiki}}\ and\ \bibinfo {author} {\bibfnamefont {K.}~\bibnamefont {Azuma}},\
  }\bibfield  {title} {\enquote {\bibinfo {title} {Quantum benchmark via an
  uncertainty product of canonical variables},}\ }\href {\doibase
  10.1103/PhysRevLett.114.140503} {\bibfield  {journal} {\bibinfo  {journal}
  {Phys. Rev. Lett.}\ }\textbf {\bibinfo {volume} {114}},\ \bibinfo {pages}
  {140503} (\bibinfo {year} {2015})}\BibitemShut {NoStop}%
\bibitem [{\citenamefont {Namiki}(2015)}]{Namiki1502}%
  \BibitemOpen
  \bibfield  {author} {\bibinfo {author} {\bibfnamefont {R.}~\bibnamefont
  {Namiki}},\ }\bibfield  {title} {\enquote {\bibinfo {title} {Amplification
  uncertainty relation for probabilistic amplifiers},}\ }\href {\doibase
  10.1103/PhysRevA.92.032326} {\bibfield  {journal} {\bibinfo  {journal} {Phys.
  Rev. A}\ }\textbf {\bibinfo {volume} {92}},\ \bibinfo {pages} {032326}
  (\bibinfo {year} {2015})}\BibitemShut {NoStop}%
\bibitem [{\citenamefont {Killoran}\ and\ \citenamefont
  {L\"utkenhaus}(2011)}]{Kil11}%
  \BibitemOpen
  \bibfield  {author} {\bibinfo {author} {\bibfnamefont {N.}~\bibnamefont
  {Killoran}}\ and\ \bibinfo {author} {\bibfnamefont {N.}~\bibnamefont
  {L\"utkenhaus}},\ }\bibfield  {title} {\enquote {\bibinfo {title} {Strong
  quantitative benchmarking of quantum optical devices},}\ }\href {\doibase
  10.1103/PhysRevA.83.052320} {\bibfield  {journal} {\bibinfo  {journal} {Phys.
  Rev. A}\ }\textbf {\bibinfo {volume} {83}},\ \bibinfo {pages} {052320}
  (\bibinfo {year} {2011})}\BibitemShut {NoStop}%
\bibitem [{\citenamefont {Killoran}\ \emph {et~al.}(2012)\citenamefont
  {Killoran}, \citenamefont {Hosseini}, \citenamefont {Buchler}, \citenamefont
  {Lam},\ and\ \citenamefont {L\"utkenhaus}}]{Kil12}%
  \BibitemOpen
  \bibfield  {author} {\bibinfo {author} {\bibfnamefont {N.}~\bibnamefont
  {Killoran}}, \bibinfo {author} {\bibfnamefont {M.}~\bibnamefont {Hosseini}},
  \bibinfo {author} {\bibfnamefont {B.~C.}\ \bibnamefont {Buchler}}, \bibinfo
  {author} {\bibfnamefont {P.~K.}\ \bibnamefont {Lam}}, \ and\ \bibinfo
  {author} {\bibfnamefont {N.}~\bibnamefont {L\"utkenhaus}},\ }\bibfield
  {title} {\enquote {\bibinfo {title} {Quantum benchmarking with realistic
  states of light},}\ }\href {\doibase 10.1103/PhysRevA.86.022331} {\bibfield
  {journal} {\bibinfo  {journal} {Phys. Rev. A}\ }\textbf {\bibinfo {volume}
  {86}},\ \bibinfo {pages} {022331} (\bibinfo {year} {2012})}\BibitemShut
  {NoStop}%
\bibitem [{\citenamefont {Khan}\ \emph {et~al.}(2013)\citenamefont {Khan},
  \citenamefont {Wittmann}, \citenamefont {Jain}, \citenamefont {Killoran},
  \citenamefont {L\"utkenhaus}, \citenamefont {Marquardt},\ and\ \citenamefont
  {Leuchs}}]{Khan13}%
  \BibitemOpen
  \bibfield  {author} {\bibinfo {author} {\bibfnamefont {I.}~\bibnamefont
  {Khan}}, \bibinfo {author} {\bibfnamefont {C.}~\bibnamefont {Wittmann}},
  \bibinfo {author} {\bibfnamefont {N.}~\bibnamefont {Jain}}, \bibinfo {author}
  {\bibfnamefont {N.}~\bibnamefont {Killoran}}, \bibinfo {author}
  {\bibfnamefont {N.}~\bibnamefont {L\"utkenhaus}}, \bibinfo {author}
  {\bibfnamefont {C.}~\bibnamefont {Marquardt}}, \ and\ \bibinfo {author}
  {\bibfnamefont {G.}~\bibnamefont {Leuchs}},\ }\bibfield  {title} {\enquote
  {\bibinfo {title} {Optimal working points for continuous-variable quantum
  channels},}\ }\href {\doibase 10.1103/PhysRevA.88.010302} {\bibfield
  {journal} {\bibinfo  {journal} {Phys. Rev. A}\ }\textbf {\bibinfo {volume}
  {88}},\ \bibinfo {pages} {010302} (\bibinfo {year} {2013})}\BibitemShut
  {NoStop}%
\bibitem [{\citenamefont {Shirokov}(2013)}]{Shirokov13}%
  \BibitemOpen
  \bibfield  {author} {\bibinfo {author} {\bibfnamefont {M.E.}\ \bibnamefont
  {Shirokov}},\ }\bibfield  {title} {\enquote {\bibinfo {title} {{Schmidt
  number and partially entanglement-breaking channels in infinite-dimensional
  quantum systems}},}\ }\href {\doibase 10.1134/S0001434613050143} {\bibfield
  {journal} {\bibinfo  {journal} {Mathematical Notes}\ }\textbf {\bibinfo
  {volume} {93}},\ \bibinfo {pages} {766--779} (\bibinfo {year}
  {2013})}\BibitemShut {NoStop}%
\bibitem [{\citenamefont {Yukawa}\ \emph {et~al.}(2008)\citenamefont {Yukawa},
  \citenamefont {Benichi},\ and\ \citenamefont {Furusawa}}]{Yukawa08}%
  \BibitemOpen
  \bibfield  {author} {\bibinfo {author} {\bibfnamefont {M.}~\bibnamefont
  {Yukawa}}, \bibinfo {author} {\bibfnamefont {H.}~\bibnamefont {Benichi}}, \
  and\ \bibinfo {author} {\bibfnamefont {A.}~\bibnamefont {Furusawa}},\
  }\bibfield  {title} {\enquote {\bibinfo {title} {High-fidelity
  continuous-variable quantum teleportation toward multistep quantum
  operations},}\ }\href {\doibase 10.1103/PhysRevA.77.022314} {\bibfield
  {journal} {\bibinfo  {journal} {Phys. Rev. A}\ }\textbf {\bibinfo {volume}
  {77}},\ \bibinfo {pages} {022314} (\bibinfo {year} {2008})}\BibitemShut
  {NoStop}%
\bibitem [{\citenamefont {Pirandola}\ \emph {et~al.}(2015)\citenamefont
  {Pirandola}, \citenamefont {Eisert}, \citenamefont {Weedbrook}, \citenamefont
  {Furusawa},\ and\ \citenamefont {Braunstein}}]{Pira15np}%
  \BibitemOpen
  \bibfield  {author} {\bibinfo {author} {\bibfnamefont {S.}~\bibnamefont
  {Pirandola}}, \bibinfo {author} {\bibfnamefont {J.}~\bibnamefont {Eisert}},
  \bibinfo {author} {\bibfnamefont {C.}~\bibnamefont {Weedbrook}}, \bibinfo
  {author} {\bibfnamefont {A.}~\bibnamefont {Furusawa}}, \ and\ \bibinfo
  {author} {\bibfnamefont {S.~L.}\ \bibnamefont {Braunstein}},\ }\bibfield
  {title} {\enquote {\bibinfo {title} {Advances in quantum teleportation},}\
  }\href@noop {} {\bibfield  {journal} {\bibinfo  {journal} {Nature Photonics}\
  }\textbf {\bibinfo {volume} {9}},\ \bibinfo {pages} {641--652} (\bibinfo
  {year} {2015})}\BibitemShut {NoStop}%
\bibitem [{\citenamefont {Ralph}\ and\ \citenamefont {Lund}()}]{Ralph09}%
  \BibitemOpen
  \bibfield  {author} {\bibinfo {author} {\bibfnamefont {T.C.}\ \bibnamefont
  {Ralph}}\ and\ \bibinfo {author} {\bibfnamefont {A.P.}\ \bibnamefont
  {Lund}},\ }\bibfield  {title} {\enquote {\bibinfo {title} {{Nondeterministic
  noiseless linear amplification of quantum systems}},}\ }\href
  {http://arxiv.org/abs/0809.0326: DOI: http://dx.doi.org/10.1063/1.3131295}
  {\bibinfo  {journal} {Quantum Communication Measurement and Computing
  Proceedings of 9th International Conference, Ed. A. Lvovsky, 155 (AIP, New
  York 2009), arXiv:0809.0326v1}\ }\BibitemShut {NoStop}%
\bibitem [{\citenamefont {Chrzanowski}\ \emph {et~al.}(2014)\citenamefont
  {Chrzanowski}, \citenamefont {Walk}, \citenamefont {Assad}, \citenamefont
  {Janousek}, \citenamefont {Hosseini}, \citenamefont {Ralph}, \citenamefont
  {Symul},\ and\ \citenamefont {Lam}}]{Chrzanowski2014}%
  \BibitemOpen
\bibfield  {journal} {  }\bibfield  {author} {\bibinfo {author} {\bibfnamefont
  {H.M.}\ \bibnamefont {Chrzanowski}}, \bibinfo {author} {\bibfnamefont
  {N.}~\bibnamefont {Walk}}, \bibinfo {author} {\bibfnamefont {S.M.}\
  \bibnamefont {Assad}}, \bibinfo {author} {\bibfnamefont {J.}~\bibnamefont
  {Janousek}}, \bibinfo {author} {\bibfnamefont {S.}~\bibnamefont {Hosseini}},
  \bibinfo {author} {\bibfnamefont {T.C.}\ \bibnamefont {Ralph}}, \bibinfo
  {author} {\bibfnamefont {T.}~\bibnamefont {Symul}}, \ and\ \bibinfo {author}
  {\bibfnamefont {P.K.}\ \bibnamefont {Lam}},\ }\bibfield  {title} {\enquote
  {\bibinfo {title} {{Measurement-based noiseless linear amplification for
  quantum communication}},}\ }\href {\doibase 10.1038/nphoton.2014.49}
  {\bibfield  {journal} {\bibinfo  {journal} {Nature Photonics}\ }\textbf
  {\bibinfo {volume} {8}},\ \bibinfo {pages} {333--338} (\bibinfo {year}
  {2014})}\BibitemShut {NoStop}%
\bibitem [{\citenamefont {Xiang}\ \emph {et~al.}(2010)\citenamefont {Xiang},
  \citenamefont {Ralph}, \citenamefont {Lund}, \citenamefont {Walk},\ and\
  \citenamefont {Pryde}}]{Xiang2010a}%
  \BibitemOpen
  \bibfield  {author} {\bibinfo {author} {\bibfnamefont {G.~Y.}\ \bibnamefont
  {Xiang}}, \bibinfo {author} {\bibfnamefont {T.~C.}\ \bibnamefont {Ralph}},
  \bibinfo {author} {\bibfnamefont {A.~P.}\ \bibnamefont {Lund}}, \bibinfo
  {author} {\bibfnamefont {N.}~\bibnamefont {Walk}}, \ and\ \bibinfo {author}
  {\bibfnamefont {G.~J.}\ \bibnamefont {Pryde}},\ }\bibfield  {title} {\enquote
  {\bibinfo {title} {{Heralded noiseless linear amplification and distillation
  of entanglement}},}\ }\href {\doibase 10.1038/nphoton.2010.35} {\bibfield
  {journal} {\bibinfo  {journal} {Nature Photonics}\ }\textbf {\bibinfo
  {volume} {4}},\ \bibinfo {pages} {316--319} (\bibinfo {year}
  {2010})}\BibitemShut {NoStop}%
\bibitem [{\citenamefont {Neergaard-Nielsen}\ \emph {et~al.}(2013)\citenamefont
  {Neergaard-Nielsen}, \citenamefont {Eto}, \citenamefont {Lee}, \citenamefont
  {Jeong},\ and\ \citenamefont {Sasaki}}]{Neergaard-Nielsen2013}%
  \BibitemOpen
  \bibfield  {author} {\bibinfo {author} {\bibfnamefont {J.S.}\ \bibnamefont
  {Neergaard-Nielsen}}, \bibinfo {author} {\bibfnamefont {Y.}~\bibnamefont
  {Eto}}, \bibinfo {author} {\bibfnamefont {C.W.}\ \bibnamefont {Lee}},
  \bibinfo {author} {\bibfnamefont {H.}~\bibnamefont {Jeong}}, \ and\ \bibinfo
  {author} {\bibfnamefont {M.}~\bibnamefont {Sasaki}},\ }\bibfield  {title}
  {\enquote {\bibinfo {title} {{Quantum tele-amplification with a
  continuous-variable superposition state}},}\ }\href {\doibase
  10.1038/nphoton.2013.101} {\bibfield  {journal} {\bibinfo  {journal} {Nature
  Photonics}\ }\textbf {\bibinfo {volume} {7}},\ \bibinfo {pages} {439--443}
  (\bibinfo {year} {2013})}\BibitemShut {NoStop}%
\bibitem [{\citenamefont {Horn}\ and\ \citenamefont {Johnson}(2007)}]{MatAnn}%
  \BibitemOpen
  \bibfield  {author} {\bibinfo {author} {\bibfnamefont {R.A.}\ \bibnamefont
  {Horn}}\ and\ \bibinfo {author} {\bibfnamefont {C.R.}\ \bibnamefont
  {Johnson}},\ }\href@noop {} {\emph {\bibinfo {title} {Matrix Analysis}}}\
  (\bibinfo  {publisher} {Cambridge, NewYork},\ \bibinfo {year}
  {2007})\BibitemShut {NoStop}%
\end{thebibliography}
%

\end{document}